\documentstyle[twocolumn,seceq]{jpsj}
\newcommand{\bc}{\begin{center}}
\newcommand{\ec}{\end{center}}
\newcommand{\be}{\begin{equation}}
\newcommand{\ee}{\end{equation}}
\newcommand{\bea}{\begin{eqnarray}}
\newcommand{\eea}{\end{eqnarray}}
\newcommand{\bse}{\begin{subequations}}
\newcommand{\ese}{\end{subequations}}
\newcommand{\bsea}{\begin{subeqnarray}}
\newcommand{\esea}{\end{subeqnarray}}

\newcommand{\detp} {\frac{{\rm d}\epsilon}{2\pi}}
\newcommand{\dtktp}{\frac{{\rm d}^{2}k}{(2\pi)^{2}}}
\newcommand{\sgn}{{\rm sgn}}
\newcommand{\msigma}{\overline{\sigma}}
\newcommand{\lsimeq}{\mbox{\raisebox{-0.6ex}{$\stackrel{<}{\sim}$}}}

\def\runtitle{
Possible Ordered States
in the 2D Extended Hubbard Model}

\def\runauthor{Masakazu {\sc Murakami}}

\title
{Possible Ordered States\\
in the 2D Extended Hubbard Model}
\author
{ 
Masakazu {\sc Murakami}\footnote{E-mail: murakami@swan.issp.u-tokyo.ac.jp}
}

\inst
{
Institute for Solid State Physics, University of Tokyo,
Roppongi, Minato-ku, Tokyo 106-8666
}

\recdate
{December 14, 1999}

\abst
{
Possible ordered states in the 2D extended Hubbard model 
with on-site ($U>0$) and nearest-neighbor ($V$) interaction 
are examined near half filling, with emphasis on the effect of finite $V$.
First, 
the phase diagram at absolute zero is determined 
in the mean field approximation. 
For $V<0$, a state where
$d_{x^{2}-y^{2}}$-wave superconductivity (dSC),
commensurate spin-density-wave (SDW) and
$\pi$-triplet pair coexist is seen to be stabilized.
Here, the importance of $\pi$-triplet pair 
on the coexistence of dSC and SDW is indicated.
This coexistent state is hampered by the phase separation (PS),
which is generally expected to occur in the presence of 
finite-range attractive interaction, but survives.
For $V>0$, a state where
commensurate charge-density-wave (CDW), SDW and ferromagnetism (FM) coexist
is seen to be stabilized.
Here, the importance of FM
on the coexistence of CDW and SDW is indicated.
Next,
in order to examine the effects of fluctuation 
on each mean field ordered state,
the renormalization group method
for the special case that the Fermi level lies just on 
the saddle points, ($\pi$,0) and (0,$\pi$), is applied.
The crucial difference from the mean field result
is that superconductivity can arise even for $U>0$ and $V\geq0$, 
where the superconducting gap symmetry is $d_{x^{2}-y^{2}}$-wave for 
$U>4V$ and $s$-wave for $U<4V$.
Finally, the possibilities that the mean field coexistent states 
survive in the presence of fluctuation are discussed.}

\kword
{2D extended Hubbard model, coexistent state,
d-wave superconductivity, s-wave superconductivity, 
spin-density wave, charge-density wave, ferromagnetism,
$\pi$-triplet pair, $\eta$-singlet pair, phase separation}

\begin{document}
\sloppy
\maketitle
\newread\epsffilein    
\newif\ifepsffileok    
\newif\ifepsfbbfound   
\newif\ifepsfverbose   
\newif\ifepsfdraft     
\newdimen\epsfxsize    
\newdimen\epsfysize    
\newdimen\epsftsize    
\newdimen\epsfrsize    
\newdimen\epsftmp      
\newdimen\pspoints     
\pspoints=1bp          
\epsfxsize=0pt         
\epsfysize=0pt         
\def\epsfbox#1{\global\def\epsfllx{72}\global\def\epsflly{72}%
   \global\def\epsfurx{540}\global\def\epsfury{720}%
   \def\lbracket{[}\def\testit{#1}\ifx\testit\lbracket
   \let\next=\epsfgetlitbb\else\let\next=\epsfnormal\fi\next{#1}}%
\def\epsfgetlitbb#1#2 #3 #4 #5]#6{\epsfgrab #2 #3 #4 #5 .\\%
   \epsfsetgraph{#6}}%
\def\epsfnormal#1{\epsfgetbb{#1}\epsfsetgraph{#1}}%
\def\epsfgetbb#1{%
%
%
\openin\epsffilein=#1
\ifeof\epsffilein\errmessage{I couldn't open #1, will ignore it}\else
%
%
   {\epsffileoktrue \chardef\other=12
    \def\do##1{\catcode`##1=\other}\dospecials \catcode`\ =10
    \loop
       \read\epsffilein to \epsffileline
       \ifeof\epsffilein\epsffileokfalse\else
%
%
          \expandafter\epsfaux\epsffileline:. \\%
       \fi
   \ifepsffileok\repeat
   \ifepsfbbfound\else
    \ifepsfverbose\message{No bounding box comment in #1; using defaults}\fi\fi
   }\closein\epsffilein\fi}%
%
%
\def\epsfclipon{\def\epsfclipstring{ clip}}%
\def\epsfclipoff{\def\epsfclipstring{\ifepsfdraft\space clip\fi}}%
\epsfclipoff
\def\epsfsetgraph#1{%
   \epsfrsize=\epsfury\pspoints
   \advance\epsfrsize by-\epsflly\pspoints
   \epsftsize=\epsfurx\pspoints
   \advance\epsftsize by-\epsfllx\pspoints
%
%
   \epsfxsize\epsfsize\epsftsize\epsfrsize
   \ifnum\epsfxsize=0 \ifnum\epsfysize=0
      \epsfxsize=\epsftsize \epsfysize=\epsfrsize
      \epsfrsize=0pt
%
%
     \else\epsftmp=\epsftsize \divide\epsftmp\epsfrsize
       \epsfxsize=\epsfysize \multiply\epsfxsize\epsftmp
       \multiply\epsftmp\epsfrsize \advance\epsftsize-\epsftmp
       \epsftmp=\epsfysize
       \loop \advance\epsftsize\epsftsize \divide\epsftmp 2
       \ifnum\epsftmp>0
          \ifnum\epsftsize<\epsfrsize\else
             \advance\epsftsize-\epsfrsize \advance\epsfxsize\epsftmp \fi
       \repeat
       \epsfrsize=0pt
     \fi
   \else \ifnum\epsfysize=0
     \epsftmp=\epsfrsize \divide\epsftmp\epsftsize
     \epsfysize=\epsfxsize \multiply\epsfysize\epsftmp   
     \multiply\epsftmp\epsftsize \advance\epsfrsize-\epsftmp
     \epsftmp=\epsfxsize
     \loop \advance\epsfrsize\epsfrsize \divide\epsftmp 2
     \ifnum\epsftmp>0
        \ifnum\epsfrsize<\epsftsize\else
           \advance\epsfrsize-\epsftsize \advance\epsfysize\epsftmp \fi
     \repeat
     \epsfrsize=0pt
    \else
     \epsfrsize=\epsfysize
    \fi
   \fi
%
%
   \ifepsfverbose\message{#1: width=\the\epsfxsize, height=\the\epsfysize}\fi
   \epsftmp=10\epsfxsize \divide\epsftmp\pspoints
   \vbox to\epsfysize{\vfil\hbox to\epsfxsize{%
      \ifnum\epsfrsize=0\relax
        \includegraphics{\ifepsfdraft}%
      \else
        \epsfrsize=10\epsfysize \divide\epsfrsize\pspoints
        \includegraphics{\ifepsfdraft}%
      \fi
      \hfil}}%
\global\epsfxsize=0pt\global\epsfysize=0pt}%
%
%
{\catcode`\%=12 \global\let\epsfpercent=
%
%
\long\def\epsfaux#1#2:#3\\{\ifx#1\epsfpercent
   \def\testit{#2}\ifx\testit\epsfbblit
      \epsfgrab #3 . . . \\%
      \epsffileokfalse
      \global\epsfbbfoundtrue
   \fi\else\ifx#1\par\else\epsffileokfalse\fi\fi}%
%
%
\def\epsfempty{}%
\def\epsfgrab #1 #2 #3 #4 #5\\{%
\global\def\epsfllx{#1}\ifx\epsfllx\epsfempty
      \epsfgrab #2 #3 #4 #5 .\\\else
   \global\def\epsflly{#2}%
   \global\def\epsfurx{#3}\global\def\epsfury{#4}\fi}%
%
%
\def\epsfsize#1#2{\epsfxsize}
%
%
\let\epsffile=\epsfbox

\section{Introduction}
In connection with the studies of
the copper oxide high-$T_{\rm c}$ superconductors with
CuO$_{2}$ planes,
the electronic states in two-dimensional systems
has been intensively studied.
Especially, the possibility of
various ordered states has been discussed.
One characteristic feature is the proximity of superconductivity 
and antiferromagnetism.
In our previous work (hereafter referred to as I),\cite{ore2and3}
we have shown in the mean field approximation
that the coexistent state with
$d$-wave superconductivity (dSC), 
commensurate spin-density-wave (SDW)
and $\pi$-triplet pair can be stabilized near half filling 
by {\em repulsive} backward scattering
('Umklapp' and 'exchange') processes
between electrons around the saddle points ($\pi$,0) and (0,$\pi$).
As we shall show later,
this model with such a particular type of interaction 
have similar features to those in a square lattice model with 
on-site {\em repulsion} $U>0$ 
and nearest-neighbor {\em attraction} $V<0$, i.e.,
an {\em extended Hubbard model}.\cite{EHM2D}
Therefore, it is interesting to examine in more detail 
the possibility of the above coexistent state
by use of this model.
At the same time, the extended Hubbard model with
both on-site and nearest-neighbor {\em repulsion} ($U>0$ and $V>0$),
is also of physical interest. 
In the 2D extended Hubbard model for $U>0$,
it has been shown based on the mean field approximation that
extended-$s$-, 
$p$- and $d$-wave superconductivity 
can arise depending on the electron density $n$ for $V<0$,
\cite{EHM2D,dag,sdd,onlyV}
and commensurate charge- and spin-density wave (CDW and SDW)
can appear at half filling $n=1$ for $V>0$.\cite{EHM2D,dag}
However, 
the property of the ground state for finite carrier doping and
the relationship among various order parameters, especially between 
dSC and SDW, have not been understood yet, 
even in the mean field approximation.
From these points of view, we will study possible 
ordered states,
especially possible coexistence of different orders,
in the 2D extended Hubbard model on a square lattice
near half filling for $U>0$ {\em and} $V\neq 0$,
with emphasis on electrons around the saddle points 
($\pi$,0) and (0,$\pi$).

In \S\ref{ehh}, the extended Hubbard model is introduced
and its relationship to our previous model used in I is referred to.
In \S\ref{mfa}, the phase diagram at absolute zero, $T=0$, 
is determined in the mean field approximation. 
In \S\ref{rg}, 
the effects of fluctuation on the mean field ordered states
are examined based on the renormalization method 
applicable only for the special case that 
the saddle points ($\pi$,0) and (0,$\pi$) lie just on the Fermi surface.

\section{Extended Hubbard Hamiltonian}\label{ehh}

The extended Hubbard Hamiltonian, $H=H_{0}+H_{U}+H_{V}$,
is written as follows:
\bse
\bea
H_{0}&=& \sum_{p\sigma}\xi_{p}c^{\dagger}_{p\sigma}c_{p\sigma},\\
\label{2DSL}
H_{U}&=&U\sum_{i}n_{i\uparrow}n_{i\downarrow}
=\frac{U}{N}\sum_{q}n_{q\uparrow}n_{-q\downarrow},\\
H_{V}&=&\frac{V}{2}\sum_{i\hat{\rho}}
n_{i}n_{i+\hat{\rho}}=\frac{1}{N}\sum_{q}V_{q}n_{q}n_{-q},
\eea
where
$\sigma$ is the spin index taking a value of $+1$ ($-1$) for 
$\uparrow$ ($\downarrow$) spin, and 
the opposite spin to $\sigma$ is denoted by $\msigma\equiv -\sigma$. 
$N$ is the total number of lattice sites,
$\xi_{p}=\epsilon_{p}-\mu$ is the one-particle energy dispersion
relative to the chemical potential $\mu$,
including nearest-neighbor- ($t$) and next-nearest-neighbor- ($t'$) hopping
integrals,
\be
\epsilon_{p}=-2t(\cos p_{x}+\cos p_{y})-4t'\cos p_{x}\cos p_{y},
\ee
$n_{q}=\sum_{\sigma}n_{q\sigma}=\sum_{k\sigma}
c^{\dagger}_{k\sigma}c_{k+q\sigma}$,
$\hat{\rho}=\pm\hat{x},\pm\hat{y}$ is the unit lattice vector and
\be
V_{q}=V(\cos q_{x}+\cos q_{y}).
\ee
\ese
The energy dispersion $\epsilon_{p}$ has two independent 
saddle points, $(\pi,0)$ and $(0,\pi)$.
In this paper, we fix $t'/t=-1/5$ with $t>0$,
in which case the Fermi surface in the absence of interaction 
approaches $(\pi,0)$ and $(0,\pi)$
as the {\em hole} doping rate, $\delta\equiv 1-n$, is increased 
from half filling, $\delta=0$.

Here, we examine the relationship between the 'g-\' ology' model used in I
and the present extended Hubbard model.\cite{oreD}
In I,
we have treated the backward scattering with large momentum transfer 
between electrons around $(\pi,0)$ and $(0,\pi)$,
i.e., 'Umklapp' ($g_{3\perp}$) and 'exchange' ($g_{1\perp}$) processes, 
and considered three types of the scattering processes, i.e.,
(1) Cooper-pair, (2) density-wave and (3) $\pi$-pair channels.\cite{ore2and3}
(Here we denote the Hamiltonian for these channels 
as $H_{1}$, $H_{2}$ and $H_{3}$.)
Especially for the repulsive case, $g_{3\perp}>0$ and $g_{1\perp}>0$,
we have shown that
the coexistent state with dSC, SDW and $\pi$-triplet pair
can be stabilized near half filling at low temperature. 
However, the above effective interaction is too simplified in that
(a) forward scattering processes are not included and
(b) the $k$-dependence of interaction is ignored.
Therefore, by transforming $H_{i}$ ($i=1,2,3$),
into real-space representation,
we will obtain a well-defined model on a square lattice.
If we keep on-site and nearest-neighbor
density-density Coulomb interaction, we obtain
\bse
\bea
H_{1}&\sim&\frac{g_{3\perp}}{2N}\biggr\{
\sum_{i}n_{i\uparrow}n_{i\downarrow}-\alpha \sum_{<ij>}\sum_{\sigma}
n_{i\sigma}n_{j\msigma}\biggr\},\\
H_{2}&\sim&\frac{g_{\perp}}{2N}\biggr\{
\sum_{i}n_{i\uparrow}n_{i\downarrow}
-\sum_{<ij>}\sum_{\sigma}n_{i\sigma}n_{j\msigma}\biggr\},\\
H_{3}&\sim&\frac{g_{1\perp}}{2N}\biggr\{
\sum_{i}n_{i\uparrow}n_{i\downarrow}-\alpha \sum_{<ij>}\sum_{\sigma}
n_{i\sigma}n_{j\msigma}\biggr\},
\eea
\label{tolattice}
\ese
where $g_{\perp}\equiv g_{3\perp}+g_{1\perp}$,
$<ij>$ stands for a bond connecting site $i$ and its nearest-neighbor
site $j$
and 
$\alpha=16/\pi^{4}\sim 0.164$.
Each $H_{i}$ is seen to describe on-site {\em repulsion} and nearest-neighbor 
{\em attraction} for $g_{3\perp},g_{1\perp}>0$. 
Therefore, 
also in the extended Hubbard model for $U>0$ and $V<0$,
the coexistent state with dSC, SDW and $\pi$-triplet pair 
is expected to be stabilized.

\section{Mean Field Analysis}\label{mfa}
First, we determine the phase diagram at absolute zero, $T=0$,
near half filling in the mean field approximation.
We fix the electron density to $n=0.9$.
With this choice of parameters, the Fermi surface of noninteracting
electrons is of the YBCO- or BSCCO-type and lies close to 
the saddle points ($\pi$,0) and (0,$\pi$),
as shown in Fig.~\ref{fspart2}.
\begin{figure}
\begin{center}
\leavevmode\epsfysize=4cm
\epsfbox{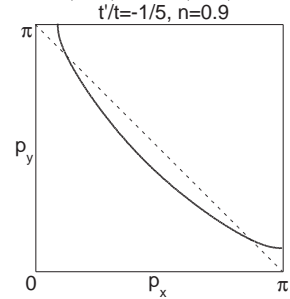}
\end{center}
\caption{The Fermi surface in the absence of interaction 
for $t'/t=-1/5$ and $n=0.9$.}
\label{fspart2}
\end{figure}

\subsection{Nearest-Neighbor Attraction $V<0$}\label{vnega}
We start with the case $V<0$.
As we saw in \S \ref{ehh},
the ordered states 
with dSC, SDW(= antiferromagnetism, AF) and 
$\pi$-triplet pair is expected.
We consider the following order parameters,
\bse
\bea
<c_{i\sigma}c_{i+\hat{\rho},\overline{\sigma}}>&\equiv& \sigma s_{\hat{\rho}}
+q_{\hat{\rho}}\cos(QR_{i}),\\
<n_{i\sigma}>&\equiv &\frac{n}{2}+\sigma m \cos(QR_{i}),
\eea
\ese
where
$Q=(\pi,\pi)$,
$s_{-\hat{\rho}}=s_{\hat{\rho}}$ and $q_{-\hat{\rho}}=q_{\hat{\rho}}$.
$s_{\hat{\rho}}$ ($q_{\hat{\rho}}$) stands for 
a spin-singlet (triplet) pair of two electrons with a total momentum $0$ ($Q$)
and total spin $S=0$ ($S=1$ and $S_{z}=0$), i.e.,
Cooper-pair ($\pi$-triplet pair),
and $m$ for the local staggered spin moment.
We take
$s_{\hat{x}}=-s_{\hat{y}}=s_{0}=real$ and 
$q_{\hat{x}}=-q_{\hat{y}}=q_{0}=real$,
i.e., consider $d_{x^{2}-y^{2}}$wave pairing,
which is favored near half filling.\cite{EHM2D}
While $s_{0}$ describes dSC,
$q_{0}$ describes an electron-pair of $p_{x}-p_{y}$-wave symmetry.
\cite{DZ,SCZhang}
This can be easily seen by writing the operator of $\pi$-triplet pair
in $k$-space,
\bse
\bea
\hat{O}_{\pi}&\equiv&\frac{1}{\sqrt{2}}
\sum_{p\sigma}w_{p}c_{-p+Q\msigma}c_{p\sigma},\\
&=&\frac{1}{\sqrt{2}}
\sum_{p\sigma}w_{\frac{Q}{2}+p}
c_{\frac{Q}{2}-p\msigma}c_{\frac{Q}{2}+p\sigma},
\eea
\ese
where $w_{p}\propto\cos p_{x}-\cos p_{y}$ 
is the $d_{x^{2}-y^{2}}$-wave factor and
$<\hat{O}_{\pi}>\propto q_{0}$.
The orbital function
as a function of the {\em relative} momentum $p$ of two electrons,
$w_{\frac{Q}{2}+p}$, is proportional to $\sin p_{x}-\sin p_{y}$.

The order parameter in the mean filed Hamiltonian are given as follows
in terms of $s_{0}$, $q_{0}$ and $m$, 
\bse
\bea
\Delta_{dSC}=-2|V|s_{0},&&
\Delta_{\pi}=-2|V|q_{0},\\
\Delta_{SDW}&=&Um,
\eea
\ese
where $\Delta_{dSC}$ and $\Delta_{\pi}$ include only $V$ because 
$s_{\hat{\rho}}$ and $q_{\hat{\rho}}$ are defined on a bond, and
$\Delta_{SDW}$ does not include $V$
because $<n_{i}>=\sum_{\sigma}<n_{i\sigma}>$ is independent of $m$.

The pure $\pi$-triplet pairing state with $\Delta_{\pi}\ne 0$ and 
$\Delta_{dSC}=\Delta_{SDW}=0$ is always energetically unfavorable 
compared with the pure dSC state with 
$\Delta_{dSC}\ne 0$ and $\Delta_{SDW}=\Delta_{\pi}=0$.
However, since the coexistence of dSC and SDW ($\Delta_{dSC}\neq 0$ 
{\em and} $\Delta_{SDW}\neq 0$)
generally results in nonzero $t_{0}$ ({\em and} nonzero $\Delta_{\pi}$ here), 
the self-consistency of mean field calculation requires the consideration
of $\pi$-triplet pair into account from the outset.
\cite{ore2and3,chmAAcoe}
The important fact 
that the coexistence of 
spin-singlet Cooper-pair and SDW always leads to 
nonzero spin-triplet pair amplitude with finite total momentum
had been recognized by Psaltakis 
{\em et al.} in a slightly different context.\cite{crete}
The close relationship among the order parameters of dSC, SDW and 
$\pi$-triplet pair is discussed in Appendix \ref{sym}.

The mean field phase diagram in the plane of $U$ and $|V|$ is shown in 
Fig.~\ref{pdvnega}.
While the dSC state is stabilized for small $U/t$,
the coexistent state with dSC, SDW and $\pi$-triplet pair is possible 
for large $U/t$,
and the phase boundary between these two states
is shown by solid line.
Although the pure $\pi$-triplet pairing state cannot be stabilized,
$\pi$-triplet pair can condensate as a result of the coexistence of 
dSC and SDW.
Since there is {\em attractive} interaction for spin-triplet channel
in the present model,
the coexistent region of dSC and SDW is widened by 
the inclusion of $\pi$-triplet pair.
We note that the Fermi surface remains in the SDW state near half filling.
As we saw in I, when SDW appears first as the temperature is lowered,
the coexistent state can be stabilized at lower temperature,
especially at $T=0$, near half filling.

\begin{figure}
\begin{center}
\leavevmode\epsfysize=5cm
\epsfbox{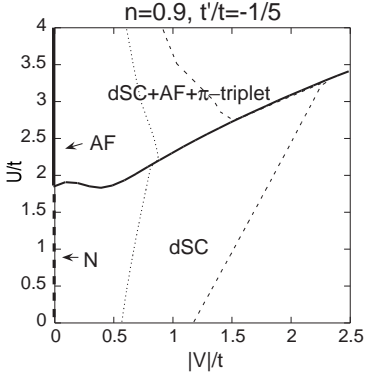}
\end{center}
\caption{The mean field phase diagram at $T=0$
for $U>0$, $V<0$, $n=0.9$ and $t'/t=-1/5$. 
Solid line stands for the boundary between
the coexistent state and dSC state in the mean field approximation.
Dotted and broken lines stand for the boundary 
in the right side of which 
the phase separation occurs in the random phase approximation
for $V'=0$ and $V'=|V|/2$, respectively,
where $V'$ is the next-nearest-neighbor density-density interaction.
$N$ stands for the normal state.}
\label{pdvnega}
\end{figure}

Generally, in the presence of finite-range attractive 
density-density interaction,
the system can be hampered by the phase separation (PS).
In order to examine the PS transition,
we calculate the charge compressibility, $\kappa$, 
given by static and uniform charge susceptibility,
in the ground state, i.e., dSC or coexistent state.
This phase boundary is determined from $\kappa^{-1}=0$.
Here we use the random phase approximation (RPA),
and take the RPA diagram, shown in Fig.~\ref{RPAdiagram1}, into account.
The explicit form of $\kappa$ in the RPA is shown in Appendix \ref{RPA}.
The PS transition line is shown in Fig.~\ref{pdvnega} by dotted line. 
It is seen that the coexistent state is severely suppressed by PS
but survives.
This PS is expected to be suppressed
if we take the long-range Coulomb repulsion into account.
Here, for simplicity,  
we consider
the {\em next}-nearest-neighbor density-density repulsion $V'>0$,
\bse
\bea
H_{V'}&=&\frac{V'}{2}\sum_{i\hat{l}}n_{i}n_{i+\hat{l}}
=\frac{1}{N}\sum_{q}V'_{q}n_{q}n_{-q},\\
V'_{q}&=&2V'\cos q_{x}\cos q_{y},
\eea
\ese
where $\hat{l}=\pm(\hat{x}+\hat{y}),\pm(\hat{x}-\hat{y})$,
and incorporate $V'$ into the RPA calculation, i.e.,
we replace $V_{q}$ with $V_{q}+V'_{q}$.
We note that this replacement, 
which does not alter the mean field equations,
can bring about the nontrivial effect on $\kappa$ in the coexistent state,
from eq. (\ref{kappaRPA}).
For $V'=|V|/2$, the PS transition line is shifted,
as shown in Fig.~\ref{pdvnega}, 
from dotted line to broken line. 
It is seen that the long-range Coulomb interaction
does lead to the suppression of the PS,
which is less prominent in the coexistent state than in the dSC state.

\begin{figure}
\begin{center}
\leavevmode\epsfysize=1.2cm
\epsfbox{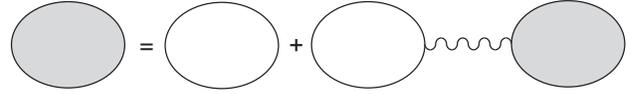}
\end{center}
\caption{The RPA diagram for charge susceptibility.}
\label{RPAdiagram1}
\end{figure}

The calculation of $\kappa$ in the RPA had been carried over by 
Micnas {\em et al.},\cite{EHM2D} {\em only} in the normal state.
Dagotto {\em et al.}\cite{dag} have shown 
based on quantum Monte Carlo (QMC) simulation that
(1) PS drastically reduces the size of the mean field dSC region
and (2) the enhancement of $d_{x^{2}-y^{2}}$ pairing correlation itself 
is not found.
This QMC result of (2) is different from the present mean field calculation.
We note that
this QMC calculation is limited to the case of half filling $n=1$ and 
relatively high temperature $T=t/6$.

The coexistence of dSC and SDW near half filling has been found 
also in the $t$-$J$ model in the slave-boson mean field 
approximation\cite{inaba} 
and by use of variational Monte Carlo calculation (VMC),\cite{Chen,giar,Himeda}
and in the repulsive Hubbard model ($V=0$) by use of VMC.\cite{giar,yamaji}
In these studies, however, $\pi$-triplet pair has not been taken into account.
The effect of $\pi$-triplet pair on the coexistence of dSC and SDW
has been recently examined by Arrachea {\em et al.}\cite{chmAAcoe}
based on a generalized Hubbard model.
In the repulsive Hubbard model, 
the nearest-neighbor hopping term is modified as the correlated one,
\bea
H_{ch}&=&-\sum_{<ij>\sigma}
\left\{c^{\dagger}_{i\sigma}c_{j\sigma}
+c^{\dagger}_{j\sigma}c_{i\sigma}\right\}\nonumber\\
&\times&
\left\{
t_{AA}(1-n_{i\msigma})(1-n_{j\msigma})
+t_{BB}n_{i\msigma}n_{j\msigma}\right.\nonumber\\
&&+t_{AB}[n_{i\msigma}(1-n_{j\msigma})+(1-n_{i\msigma})n_{j\msigma}]
\left.\right\},
\eea
where three hopping integrals, $t_{AA}$, $t_{BB}$ and $t_{AB}$,
incorporate many-body effects into one-particle hopping processes
phenomenologically.
$t_{AA}$ and $t_{BB}$ do not change the number of doubly occupied sites,
and $t_{AB}$ does, as shown in Fig.~\ref{chopping}.      
It is to be noted that $H_{ch}$ can be rewritten as
\bse
\bea
H_{ch}&=&\sum_{<ij>\sigma}
\left\{c^{\dagger}_{i\sigma}c_{j\sigma}
+c^{\dagger}_{i\sigma}c_{j\sigma}\right\}\nonumber\\
&\times&
\left\{-t+
t_{2}(n_{i\msigma}+n_{j\msigma})
+t_{3}n_{i\msigma}n_{j\msigma}\right\},
\eea
where 
\be
t\equiv t_{AA},\:\:t_{2}\equiv t_{AA}-t_{AB},\:\:
t_{3}\equiv 2t_{AB}-t_{AA}-t_{BB}.
\ee
\ese
The $t_{2}$ term can be also deduced
from the bare Coulomb interaction\cite{BCR1D,BCRHS,PH1D}
or by including the effects of phonon 
in the antiadiabatic approximation $M\rightarrow 0$ 
(where $M$ is the phonon mass),\cite{BCAAL}
and the $t_{3}$ term describes the three-body interaction.
Arrachea {\em et al.} have shown
in the mean field approximation for 
$t_{AB}>t_{AA}=t_{BB}$ (i.e., $t_{2}<0$ and $t_{3}=-2t_{2}>0$)
and $t'=0$ that 
the coexistence of dSC and SDW is possible but prevented
by $\pi$-triplet pair
(and ruled out for large $U$), 
due to {\em repulsive} spin-triplet pairing interaction.
Hence, the effect of $\pi$-triplet pair on the coexistence of 
dSC and SDW is different from that in the present extended Hubbard model.

\begin{figure}
\begin{center}
\leavevmode\epsfysize=4.5cm
\epsfbox{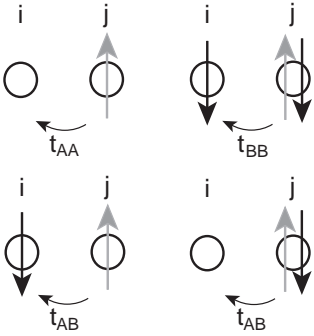}
\end{center}
\caption{The correlated hopping processes.}
\label{chopping}
\end{figure}

\subsection{Nearest-Neighbor Repulsion $V>0$}\label{vposi}
Next we treat the case $V>0$.
In this case, 
not only charge- and spin-density-wave states (CDW and SDW),
but also orbital antiferromagnetic (OAF) and spin nematic (SN) states, 
in which the local staggered currents of charge and spin 
circulate, respectively,\cite{fluxphase,fsinstability,OAF,SN}
are expected.\cite{CG}
Here we include ferromagnetism (FM) for the reason 
as we shall describe later.
The order parameters are 
\bse
\be
<n_{i\sigma}>\equiv \frac{n}{2}+\sigma f+
(p+\sigma m) \cos(QR_{i}),\label{dw}
\ee
\be
<c^{\dagger}_{i\sigma}c_{i+\hat{\rho},\sigma}>\equiv 
(g_{\hat{\rho}}+\sigma l_{\hat{\rho}})\cos(QR_{i}),\label{OAFSN2D}
\ee
\ese
where $f$, $p$ and $m$ are real,
$g_{-\hat{\rho}}=-g^{\ast}_{\hat{\rho}}$ and 
$l_{-\hat{\rho}}=-l^{\ast}_{\hat{\rho}}$.
$f$, $p$ and $m$ describes FM, CDW and SDW, respectively. 
We take $g_{\hat{\rho}}$
and $l_{\hat{\rho}}$ to be of $d_{x^{2}-y^{2}}$wave symmetry,
which is favored near half filling.\cite{CG}
In this case, $g_{\hat{\rho}}$
and $l_{\hat{\rho}}$ become pure imaginary,
$g_{\pm\hat{x}}=-g_{\pm\hat{y}}={\rm i}g$ and
$l_{\pm\hat{x}}=-l_{\pm\hat{y}}={\rm i}l$,
where $g$ and $l$ are real.
The states with $g\ne 0$ and $l\ne 0$ are called as OAF and SN ones,
\cite{fluxphase,fsinstability,OAF,SN}
respectively,
in which there exist the staggered 
local currents of charge and spin
on a bond ($i,i+\hat{\rho}$),
\bse
\bea
<j^{c}_{i,i\pm\hat{x}}>=-<j^{c}_{i,i\pm\hat{y}}>&\propto &g\cos(QR_{i}),\\
<j^{s}_{i,i\pm\hat{x}}>=-<j^{s}_{i,i\pm\hat{y}}>&\propto &l\cos(QR_{i}),
\eea
\ese
where
\be
j^{\nu}_{i,i+\hat{\rho}}\equiv {\rm i}\sum_{\sigma}v_{\nu}
(c^{\dagger}_{i\sigma}c_{i+\hat{\rho},\sigma}-
c^{\dagger}_{i+\hat{\rho},\sigma}c_{i\sigma}),
\ee
and $v_{c}=1$ and $v_{s}=\sigma$.
In the OAF (SN) state, the local staggered current of charge (spin) 
circulates around the plaquettes,
as schematically shown in Fig.~\ref{oafsn},
and the bond-ordered wave (BOW) does not exist, i.e., 
$<c^{\dagger}_{i\sigma}c_{i+\hat{\rho},\sigma}+
c^{\dagger}_{i+\hat{\rho},\sigma}c_{i\sigma}>\equiv 0$.

\begin{figure}
\begin{center}
\leavevmode\epsfysize=2.5cm
\epsfbox{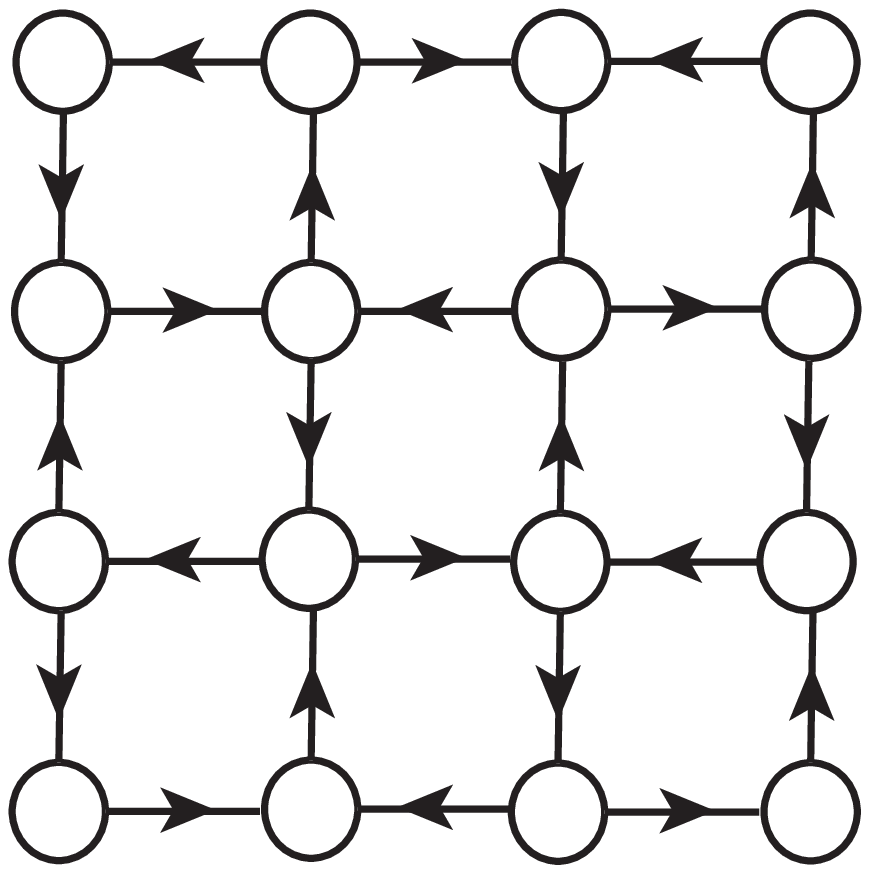}
\end{center}
\caption{The OAF (SN) state in which the local staggered 
charge (spin) current circulates around the plaquettes.}
\label{oafsn}
\end{figure}

The order parameters in the mean field Hamiltonian
are given as follows in terms of $f$, $p$, $m$, $g$ and $l$,
\begin{subequations}
\bea
\Delta_{FM}&=&Uf,\\
\Delta_{CDW}=(8V-U)p,&&\Delta_{SDW}=Um,\\
\Delta_{OAF}=2Vg,&&\Delta_{SN}=2Vl,
\eea
\end{subequations}
where $\Delta_{OAF}$ and $\Delta_{SN}$ include only $V$
because $g_{\hat{\rho}}$ and $l_{\hat{\rho}}$ are defined on a bond,
and $\Delta_{FM}$ does not include $V$
because $<n_{i}>=\sum_{\sigma}<n_{i\sigma}>$ is independent of $f$.
The order parameters of CDW, SDW, OAF and SN are closely related to 
each other, which is shown in Appendix \ref{sym}.

The mean field phase diagram in the plane of $U$ and $V$ is shown 
in Fig.~\ref{pdvposi}.
We have found that a coexistent solution
with nonzero $\Delta_{CDW}$, $\Delta_{SDW}$
{\em and} $\Delta_{FM}$
can be stabilized for $n\neq 1$.
This state is {\em ferrimagnetic}, as shown in Fig.~\ref{ferri}.
We note that the coexistence of CDW and SDW 
($\Delta_{CDW}\neq 0$ {\em and} $\Delta_{SDW}\neq 0$)
generally results
in nonzero $f$ ({\em and} nonzero $\Delta_{FM}$ here),
which had been indicated by Dzyaloshinski\u\i\cite{scvHs}
based on a qualitative symmetry analysis.
This is the reason why we take FM into account from the outset.
With the present choice of parameters,
pure FM state with only $\Delta_{FM}\neq 0$ cannot be stabilized, but
FM can arise as a result of the coexistence of CDW and SDW.
In the present case, the coexistent region of CDW and SDW
is widened by the inclusion of FM.
We note that the Fermi surface remains in the CDW or SDW state 
near half filling.

It is to be noted that neither OAF nor SN can been stabilized solely
in the mean field approximation,
{\em independent} of $t'$, $U$, $V$, $T$ and $n$.
This conclusion is contrary to that of 
Chattopadhyay {\em et al.}\cite{CG} 
that pure OAF or SN state has lower ground-state energy
than pure CDW or SDW state for the half-filled case
by introducing finite $t'$.
Moreover, a state where
local-current (OAF or SN) and density-wave (CDW or SDW) coexist 
cannot be also stabilized.
\begin{figure}
\begin{center}
\leavevmode\epsfysize=5cm
\epsfbox{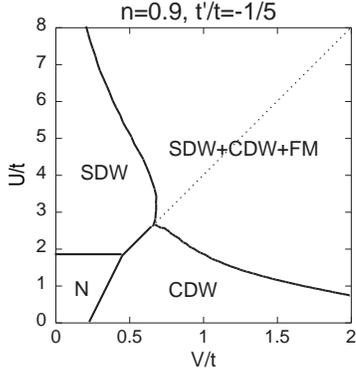}
\end{center}
\caption{The mean field phase diagram at $T=0$ for $U>0$, $V>0$, $n=0.9$ 
and $t'/t=-1/5$.}
\label{pdvposi}
\end{figure}
\begin{figure}
\begin{center}
\leavevmode\epsfysize=3cm
\epsfbox{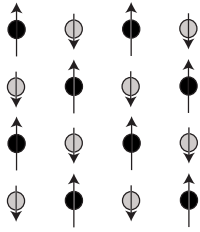}
\end{center}
\caption{The ferrimagnetic coexistent state.
Each lattice site is shaded according to electron density.
The length of each arrow is proportional to 
the magnitude of local spin moment.
Lattice sites with larger local spin moment have higher electron density.}
\label{ferri}
\end{figure}

For $U/t=4.0$,
the $V$ dependences of $\Delta_{CDW}$, $\Delta_{SDW}$,
$\Delta_{FM}$ and
the difference between the energy of the 
pure state (CDW for $U<4V$ or SDW for $U>4V$), $E_{p}$, and that of 
the coexistent state (CDW+SDW+FM), $E_{c}$, 
are shown in Fig.~\ref{cdwsdw}.
In the coexistent state with CDW, SDW and FM, 
$|\Delta_{SDW}|$ is larger than $|\Delta_{CDW}|$
for $U>4V$ and vice versa for $U<4V$, and
the first-order phase transition occurs at $U=4V$.
For fixed $U$, $|\Delta_{FM}|$ rapidly saturates as a function of $V$.
It is seen that the energy gain in the coexistent state is very small.
\begin{fullfigure}
\begin{center}
\leavevmode\epsfysize=5cm
\epsfbox{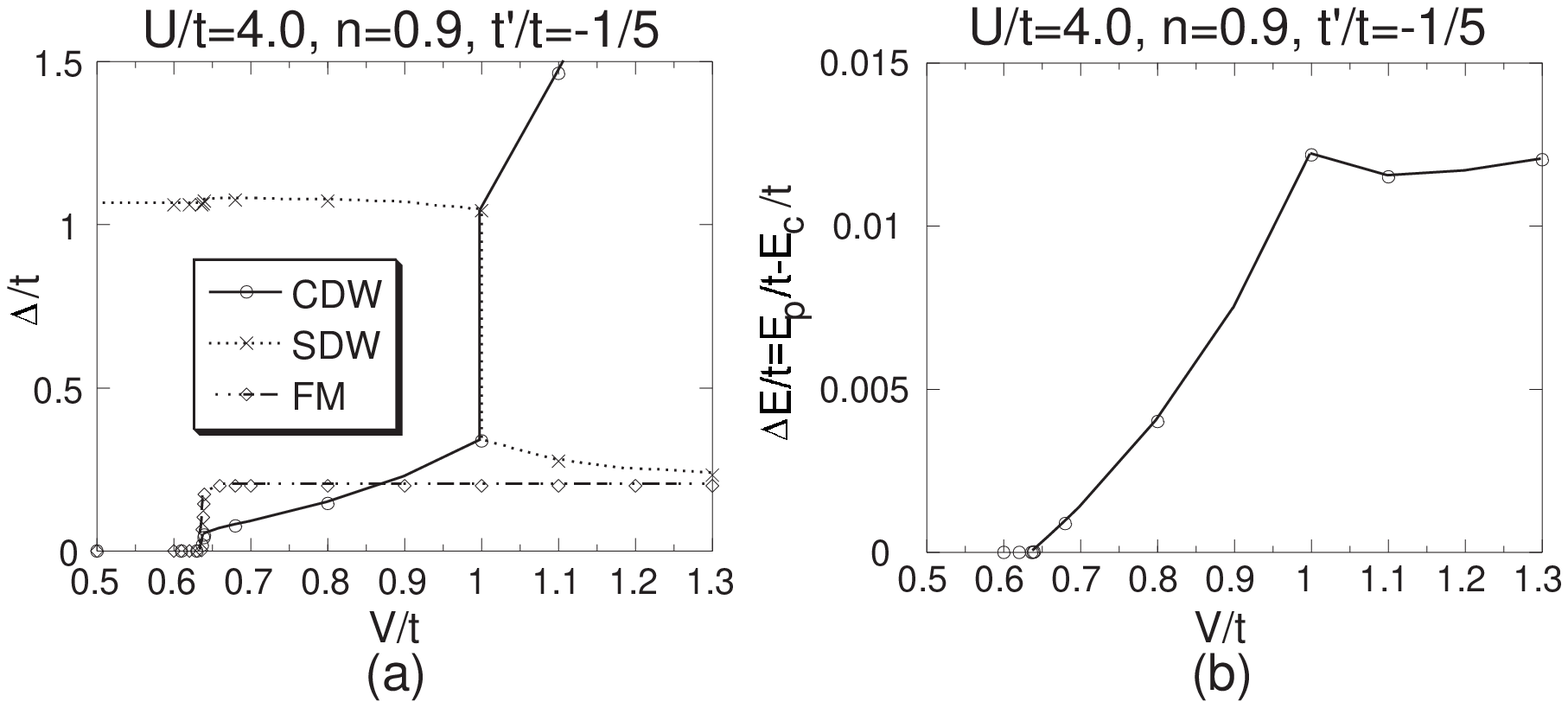}
\end{center}
\caption{The $V$ dependences of (a) $\Delta_{CDW}$, $\Delta_{SDW}$ 
$\Delta_{FM}$ and
(b)the energy difference $E_{p}-E_{c}$ (scaled by $t$)
between the pure state (CDW or SDW) and the coexistent state (CDW+SDW+FM)
for $U/t=4.0$, $n=0.9$ and $t'/t=-1/5$.}
\label{cdwsdw}
\end{fullfigure}

The energy dispersion in the ferrimagnetic coexistent state is
shown in Fig.~\ref{cdwsdwdisp}.
There are four energy bands, and 
the Fermi surface remains as in the CDW or SDW state.
However, the lower band of electrons with majority spin 
(up spin for $\Delta_{FM}>0$) is fully occupied and 
the Fermi level crosses only the lower band of electrons 
with minority spin (down spin for $\Delta_{FM}>0$).
Therefore, this coexistent state is {\em half metallic}. 
\begin{figure}
\begin{center}
\leavevmode\epsfysize=5cm
\epsfbox{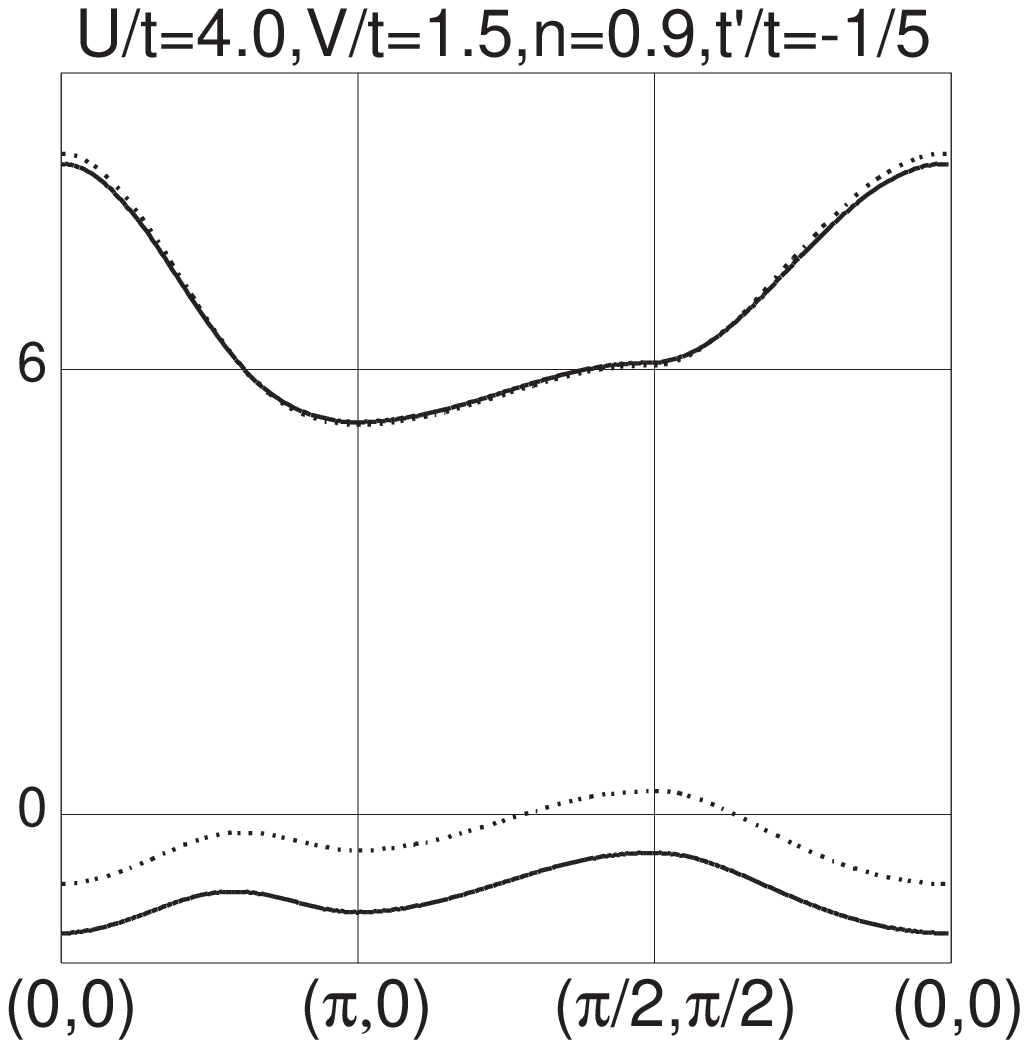}
\end{center}
\caption{The energy dispersion relative to the Fermi level
in the coexistent state with CDW, SDW and FM
for $U/t=4.0$, $V/t=1.5$, $n=0.9$ and $t'/t=-1/5$.
Full (dotted) lines stand for that of electrons with up (down) spin, 
respectively. The lower band of electrons with up spin is fully occupied.}
\label{cdwsdwdisp}
\end{figure}

The coexistence of CDW and SDW with {\em same} wave vectors
has also been found in a 1D modified Hubbard model for a quarter-filled band
in the mean field approximation.\cite{1D2kf}
In this coexistent state,
the wave vector of charge and spin density, $q$,
and the phase difference between CDW and SDW, $\Delta\theta$, 
are equal to $2k_{F}\equiv \pi/2$ and $\pi/2$, respectively,
and the magnitude of local spin moment is equal at each site,
as shown in Fig.~\ref{kogata}.
On the other hand, in our coexistent state,
$q=Q\equiv (\pi,\pi)$ and $\Delta\theta =0$,
and the magnitude of local spin moment is different 
at each sublattice, as shown in Fig.~\ref{ferri}.
This {\em ferrimagnetic} coexistent state is 
the 2D version of that found in the 3D Hubbard model ($V=0$)
which had been denoted as the special ferrimagnetic (S.F.) state. \cite{penn} 
In the present 2D case, this coexistent state for $V=0$ 
can be stabilized for $12\lsimeq U/t\lsimeq 14$ 
(not shown in Fig.~\ref{pdvposi}).

\begin{figure}
\begin{center}
\leavevmode\epsfysize=0.7cm
\epsfbox{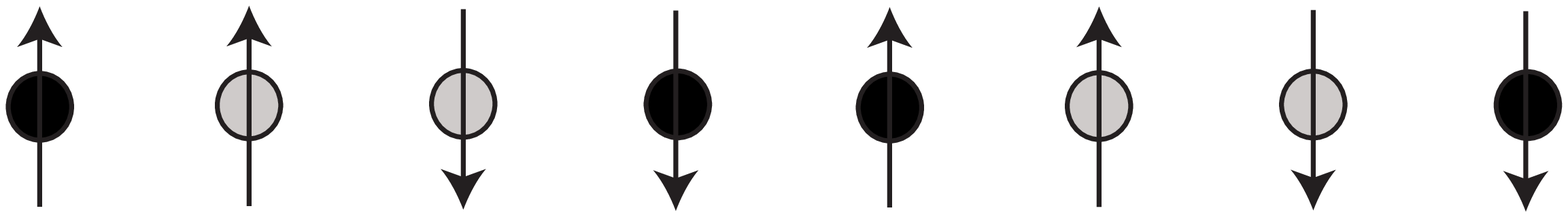}
\end{center}
\caption{The coexistent state with $2k_{F}$ CDW and $2k_{F}$ SDW
found in a quarter-filled 1D modified Hubbard model.
\protect\cite{1D2kf}
Each lattice site is shaded according to electron density.
The length of each arrow, proportional to 
the magnitude of local spin moment,
is equal at each site.}
\label{kogata}
\end{figure}

\section{Renormalization Group Analysis}\label{rg}
In the last section,
we examined possible ordered states for $U>0$ and $V\ne 0$
in the mean field approximation.
In this section,
we examine the effects of fluctuation on these ordered states
which are not taken into account in the mean field calculation.
As a theoretical treatment beyond the mean field level,
we adopt the renormalization group (RG) method for the saddle points
which has been applied to the Hubbard model ($V=0$),
\cite{Schulz,Lederer,scvHs,vHsRG,ttu,ttu2,FurukawaRice}
and determine the most dominant correlation in the normal state.
We also discuss the possibility of the coexistent states
beyond the mean field approximation.

\subsection{Saddle Point Singularity}

We consider the {\em special} case 
where the Fermi level in the absence of interaction lies 
{\em just} on the saddle points $Q_{A}\equiv (\pi,0)$ 
and $Q_{B}\equiv (0,\pi)$, i.e., $\mu=4t'$ ($n\sim 0.83$),
and focus on electrons at these two saddle points
{\em on the Fermi surface}, just as two Fermi points in 1D electron systems.
The Fermi surface is shown in Fig.~\ref{juston}.

\begin{figure}
\begin{center}
\leavevmode\epsfysize=4cm
\epsfbox{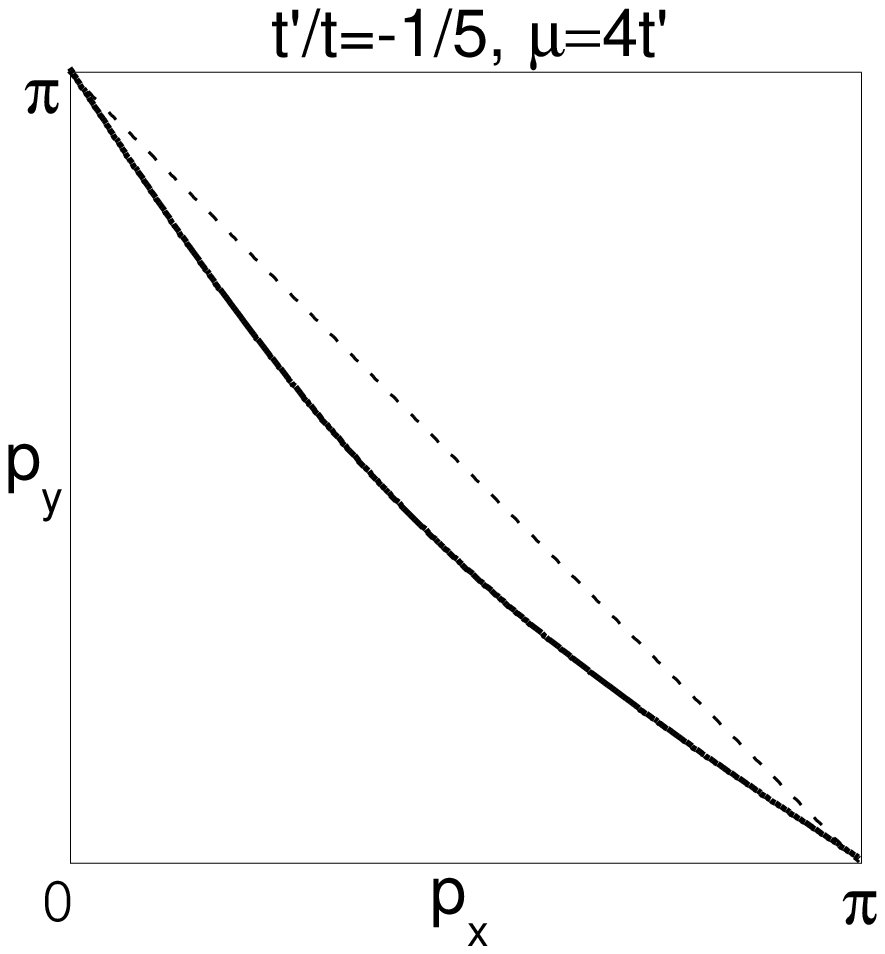}
\end{center}
\caption{The Fermi surface in the absence of interaction 
for $t'/t=-1/5$ and $\mu=4t'$, in which case $n\sim 0.83$.}
\label{juston}
\end{figure}

First, we examine the behavior of the following 
particle-particle (K) and particle-hole (P) correlation functions,
\bse
\bea
K_{\alpha\alpha'}&=&\lim_{q\rightarrow 0}\int_{k,\epsilon}
G_{\alpha}(k,\epsilon)G_{\alpha'}(-k+q,-\epsilon),\label{ppcf}\\
P_{\alpha\alpha'}&=&\lim_{q\rightarrow 0}\int_{k,\epsilon}
G_{\alpha}(k,\epsilon)G_{\alpha'}(k+q,\epsilon)\label{phcf},
\eea
\ese
for $\alpha,\alpha'=A,B$, where
\be
G_{\alpha}(k,\epsilon)\equiv\frac{1}{{\rm i}\epsilon
-\xi_{Q_{\alpha}+k}}, 
\ee
is the one-particle Green function in the absence of interaction 
for electrons near the saddle point $Q_{\alpha}$ and
\be
\int_{k,\epsilon}\equiv \int_{|k|<k_{c}} 
\dtktp\int\detp,
\ee
where the cutoff around the saddle points, $k_{c}$, is introduced.
We note that
\bse
\be
K_{1}\equiv K_{AA}=K_{BB},\:\:K_{2}\equiv K_{AB}=K_{BA},\label{K1K2}
\ee
stand for Cooper- and $\pi$-pair correlation, respectively, and 
\be
P_{1}\equiv P_{AA}=P_{BB},\:\:P_{2}\equiv P_{AB}=P_{BA},\label{P1P2}
\ee
\ese
for uniform and staggered density-density correlation, respectively.
For $\mu=4t'$, these correlation functions are logarithmically divergent,
\bse
\be
K_{1}\sim \frac{c}{8\pi^{2}t}\log^{2}\frac{E_{c}}{\omega}, \:\:
P_{1}\sim -\frac{c}{4\pi^{2}t}\log\frac{E_{c}}{\omega},
\ee
\be
K_{2}\sim\left\{
\begin{array}{cc} 
\frac{c''}{4\pi^{2}t}\log\frac{E_{c}}{\omega} 
& \mbox{for } \omega \ll rE_{c},\\
-P_{1} & \mbox{for } \omega \gg rE_{c},
\end{array}\right.
\ee
\be
P_{2}\sim 
\left\{
\begin{array}{cc}
-\frac{c'}{4\pi^{2}t}\log\frac{E_{c}}{\omega} 
& \mbox{for } \omega \ll rE_{c},\\
-K_{1} & \mbox{for } \omega \gg rE_{c},
\end{array}
\right.
\ee
\label{bubbles2}
\ese
where $E_{c}>0$ and $\omega>0$ ($\omega\ll E_{c}$)
are the ultraviolet and infrared energy cutoff, respectively,
\bse
\bea
c&\equiv &\frac{1}{\sqrt{1-4r^{2}}},\\
c'&\equiv&\log\frac{1+\sqrt{1-4r^{2}}}{2r},\\
c''&\equiv&\frac{1}{2r}\arctan(\frac{2r}{\sqrt{1-4r^{2}}}),
\eea
\label{coefc}
\ese
and $r\equiv |t'|/t$.
$c$, $c'$ and $c''$ as a function of $r$ are shown in Fig.~\ref{coef}.
Especially for small $r$,
\be
c,c''\sim 1,\:\: c'\sim -\log r.\label{smallt'}
\ee
For $r>r_{c}\sim 0.276$, $c>c'$ and $P_{1}$ is more divergent than
$P_{2}$ for $\omega\rightarrow 0$.
We note $c''<$max$\{c,c'\}$, i.e., $\pi$-pair susceptibility $K_{2}$
is always less divergent than particle-hole susceptibility.
For $t'/t=-0.2$, $c''<c<c'$ and they are comparable in magnitude.

\begin{figure}
\begin{center}
\leavevmode\epsfysize=4.5cm
\epsfbox{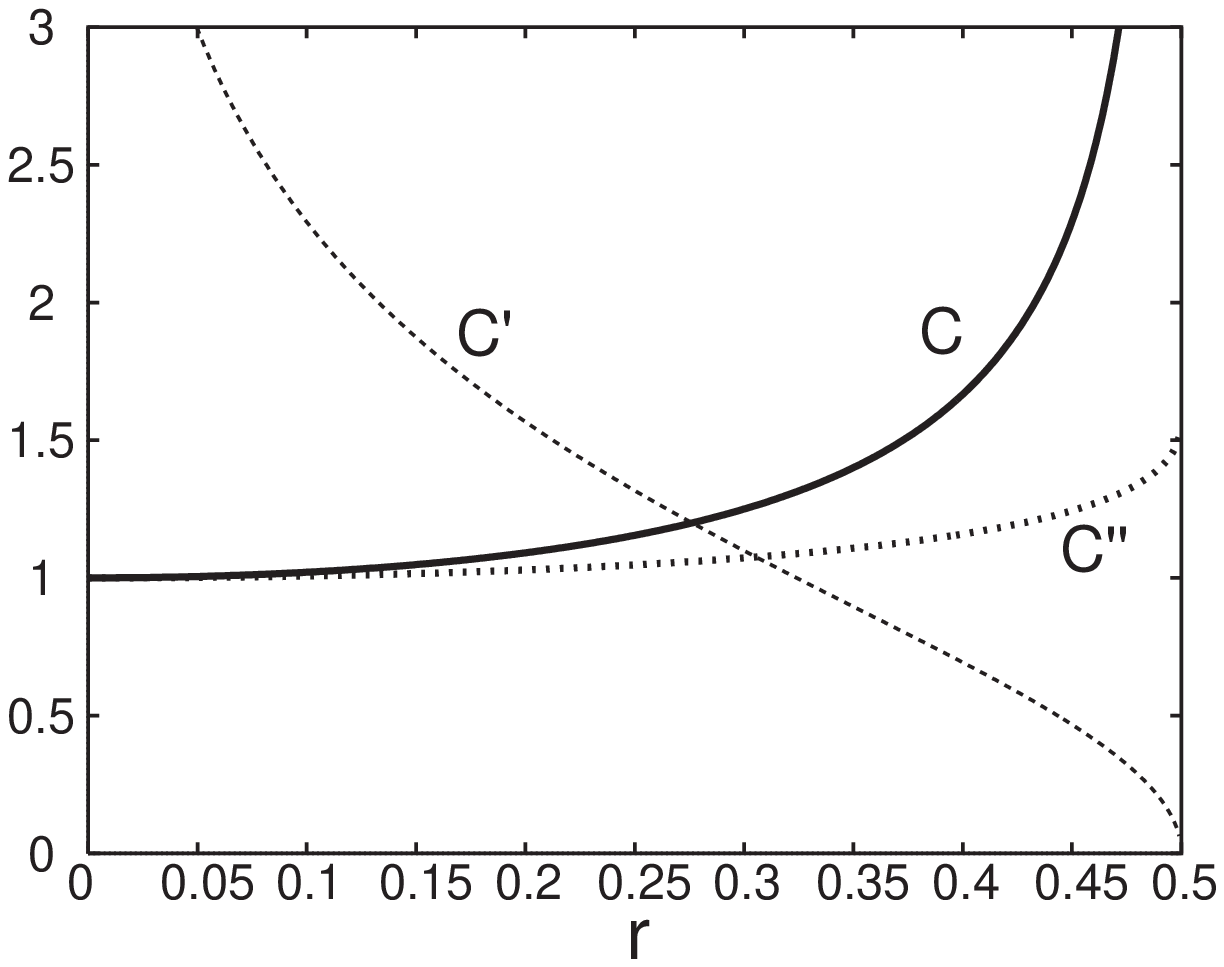}
\end{center}
\caption{$c$, $c'$ and $c''$ as a function of $r$.}
\label{coef}
\end{figure}

\subsection{Renormalization Group Method for the Saddle Points}\label{RGtech}

In the last subsection,
we saw that the saddle points on the Fermi surface lead to 
logarithmic divergence of particle-particle and particle-hole  
correlation functions.
This implies that
the fluctuation effect becomes strong.
In the RG approach,
we assume that the single renormalization group variable 
$x\equiv\log\frac{E_{c}}{\omega}$ determine the behavior of the system.
The increase of $x$ represents renormalization towards lower energy scale.
For simplicity, we neglect
(1) the deformation of the Fermi surface 
by interaction and
(2) $k$-dependence of interaction for small $|k|<k_{c}$,
i.e., we consider only eight coupling constants, 
$g_{is}$ ($i=1,2,3,4$ and $s=\perp,\parallel$).
This interaction of the g-\' ology type is shown in Fig.~\ref{gology}.
$g_{1}$ and $g_{3}$
($g_{2}$ and $g_{4}$)
stand for the backward (forward) scattering processes with large (small)
momentum transfer, respectively. 
Especially, $g_{1}$ and $g_{3}$
describe 'exchange' and 'Umklapp' processes.
In I, only $g_{1\perp}$ and $g_{3\perp}$ were treated
and taken to be momentum-independent 
all over the magnetic Brillouin zone.\cite{ore2and3}

\begin{figure}
\begin{center}
\leavevmode\epsfysize=4cm
\epsfbox{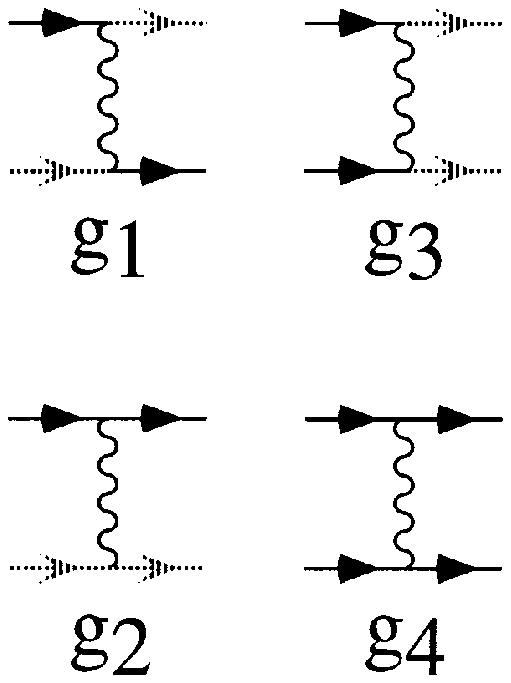}
\end{center}
\caption{The scattering processes. 
Solid and dashed lines stand for electrons 
near $Q_{A}=(\pi,0)$ and $Q_{B}=(0,\pi)$, respectively.}
\label{gology}
\end{figure}

We start with
the renormalization of the couplings in the one-loop approximation.
One-loop diagrams are shown in Fig.~\ref{G1loop}.
The scaling equations are 
\bse
\bea
\dot{g_{1\perp}}&=&
-2g_{1\perp}g_{2\perp}\dot{K_{2}}-2g_{1\perp}g_{4\perp}\dot{P_{1}}
\nonumber\\&&
+2g_{1\perp}(g_{1\parallel}-g_{2\parallel})\dot{P_{2}},\\
\dot{g_{1\parallel}}&=&
-2g_{1\parallel}g_{2\parallel}\dot{K_{2}}-2g_{1\parallel}g_{4\parallel}
\dot{P_{1}}\nonumber\\&&
+[2g_{1\parallel}(g_{1\parallel}-g_{2\parallel})
+(g_{1\perp}^{2}-g_{1\parallel}^{2})\nonumber\\&&
\:\:\:\:\:\:+(g_{3\perp}^{2}-g_{3\parallel}^{2})]\dot{P_{2}},\\
\dot{g_{2\perp}}&=&
-(g_{1\perp}^{2}+g_{2\perp}^{2})\dot{K_{2}}-2g_{4\perp}(g_{1\parallel}-
g_{2\parallel})\dot{P_{1}}
\nonumber\\&&
-(g_{2\perp}^{2}+g_{3\perp}^{2})\dot{P_{2}},\\
\dot{g_{2\parallel}}&=&
-(g_{1\parallel}^{2}+g_{2\parallel}^{2})\dot{K_{2}}
-2(g_{4\parallel}g_{1\parallel}-g_{4\perp}g_{2\perp})\dot{P_{1}}
\nonumber\\&&
-(g_{2\parallel}^{2}+g_{3\parallel}^{2})\dot{P_{2}},\\
\dot{g_{3\perp}}&=&-2g_{3\perp}g_{4\perp}\dot{K_{1}}
-2g_{3\perp}(g_{2\perp}+g_{2\parallel}-g_{1\parallel})\dot{P_{2}},
\label{g3perp}\\
\dot{g_{3\parallel}}&=&-2g_{3\parallel}g_{4\parallel}\dot{K_{1}}
-2(2g_{3\parallel}g_{2\parallel}-g_{3\perp}g_{1\perp})\dot{P_{2}},\\
\dot{g_{4\perp}}&=&-(g_{3\perp}^{2}+g_{4\perp}^{2})\dot{K_{1}}
\nonumber\\&&-[g_{1\perp}^{2}+g_{4\perp}^{2}+
2g_{2\perp}(g_{1\parallel}-g_{2\parallel})]\dot{P_{1}},\\
\dot{g_{4\parallel}}&=&-(g_{3\parallel}^{2}+g_{4\parallel}^{2})\dot{K_{1}}
\nonumber\\&&-[g_{1\parallel}^{2}+(2g_{4\parallel}^{2}-g_{4\perp}^{2})
+2g_{2\parallel}(g_{1\parallel}-g_{2\parallel})
\nonumber\\&&\:\:\:\:\:\:
-(g_{2\perp}^{2}-g_{2\parallel}^{2})]\dot{P_{1}},
\eea
\label{intflow}
\ese
which are to be solved with the initial conditions 
$g_{is}(x=x_{i})=g^{0}_{is}$ ( $\dot{}$ $\equiv $d/d$x$),
where $g^{0}_{is}$ are the bare coupling constants.
For example, $g_{3\perp}$ has the following form
to one loop order,
\be
g_{3\perp}=-2g^{0}_{3\perp}g^{0}_{4\perp}K_{1}
-2g^{0}_{3\perp}(g^{0}_{2\perp}+g^{0}_{2\parallel}-g^{0}_{1\parallel})
P_{2}.
\ee
By differentiating this equation by $x$ and
replace $g^{0}_{is}$ by $g_{is}$, i.e.,
the bare coupling constants by the renormalized ones,
we obtain the scaling equation eq. (\ref{g3perp}).\\
If we take $g^{0}_{i\perp}=g^{0}_{i\parallel}$ as the initial conditions, 
the relation $g_{i\perp}=g_{i\parallel}$ holds all through the flow.
Therefore, the above scaling equations are simplified 
as follows,\cite{FurukawaRice}
\bse
\bea
\dot{g_{1}}&=&-2g_{1}g_{2}\dot{K_{2}}-2g_{1}g_{4}\dot{P_{1}}
-2g_{1}(g_{2}-g_{1})\dot{P_{2}},\\
\dot{g_{2}}&=&-(g_{1}^{2}+g_{2}^{2})\dot{K_{2}}
+2g_{4}(g_{2}-g_{1})\dot{P_{1}}\nonumber\\&&
-(g_{2}^{2}+g_{3}^{2})\dot{P_{2}},\\
\dot{g_{3}}&=&-2g_{3}g_{4}\dot{K_{1}}-2g_{3}(2g_{2}-g_{1})\dot{P_{2}},\\
\dot{g_{4}}&=&-(g_{3}^{2}+g_{4}^{2})\dot{K_{1}}\nonumber\\&&
-[g_{1}^{2}+g_{4}^{2}+
-2g_{2}(g_{2}-g_{1})]\dot{P_{1}}.
\eea
\label{intflow2}
\ese
where $g_{i}\equiv g_{i\perp}=g_{i\parallel}$.

The divergence of $g_{is}(x)$ at a {\em finite} $x$  
indicates the existence of the strong coupling fixed point,
i.e., signals the development of an ordered state,
at {\em finite} energy scale or {\em finite} temperature.
(Strictly speaking, this {\em finite} onset temperature 
is an artifact of the present approximation in the 2D systems,
and should be interpreted as a crossover temperature, or a 
critical temperature when finite three-dimensionality is assumed.)
The properties of this strong coupling fixed point
can be obtained qualitatively from various response functions.
The response functions in the one-loop approximation 
are obtained from one-loop diagrams shown in Fig.~\ref{R1loop}.
The response function,
\be
R_{\nu}=\int_{0}^{\beta}{\rm d}\tau {\rm e}^{{\rm i}0\tau}\cdot
\frac{1}{N}<T_{\tau}\hat{O}^{\dagger}_{\nu}(\tau)\hat{O}_{\nu}>,
\ee
where $\nu$ stands for the kind of correlation ($\nu=$dSC, SDW, $\cdots$), 
has the following form to one-loop order,
\be
R_{\nu}=R_{\nu}^{0}+\frac{1}{4}g^{0}_{\nu}(R_{\nu}^{0})^{2},
\ee
where $g^{0}_{\nu}$ is the coupling constant (linear combination of 
$g^{0}_{is}$) and $R_{\nu}^{0}$ is the simple bubble.
If we differentiate this equation by $x$ and
replace $g_{\nu}^{0}$, $R_{\nu}^{0}$ by the renormalized ones,
$g_{\nu}$ and $R_{\nu}$,
we obtain
\be
\dot{R_{\nu}}=\dot{R_{\nu}^{0}}
\left\{1+\frac{1}{2}g_{\nu}R_{\nu}\right\}.\label{Rnflow}
\ee
This equation is to be solved
with the initial condition that $R_{\nu}(x=x_{i})\sim 0$.
Since $\dot{R_{\nu}}^{0}$ is positive,
$R_{\nu}$ can be divergent for $g_{\nu}>0$ and are suppressed to zero for
$g_{\nu}<0$.
In this paper,
we consider the response functions shown in Fig~\ref{gn},
where
$\hat{O}_{sSC}$, $\hat{O}_{\eta}$ and $\hat{O}_{PS}$ 
stand for  $s$-wave Cooper-pair,
$\eta$-singlet pair with a total momentum $Q$ 
and total spin $S=0$,\cite{Eta1,Eta2} 
and uniform charge density, respectively.
The most divergent $R_{PS}$ is interpreted
to describe the phase separation (PS).
The relationship among these order parameters is discussed in
Appendix \ref{sym}.
We note that each correlation is treated independently 
in the above procedure.
Therefore,
we can determine the most dominant susceptibility in the normal state, 
and cannot assess the coexistence of different orders.

\begin{figure}
\begin{center}
\leavevmode\epsfysize=5cm
\epsfbox{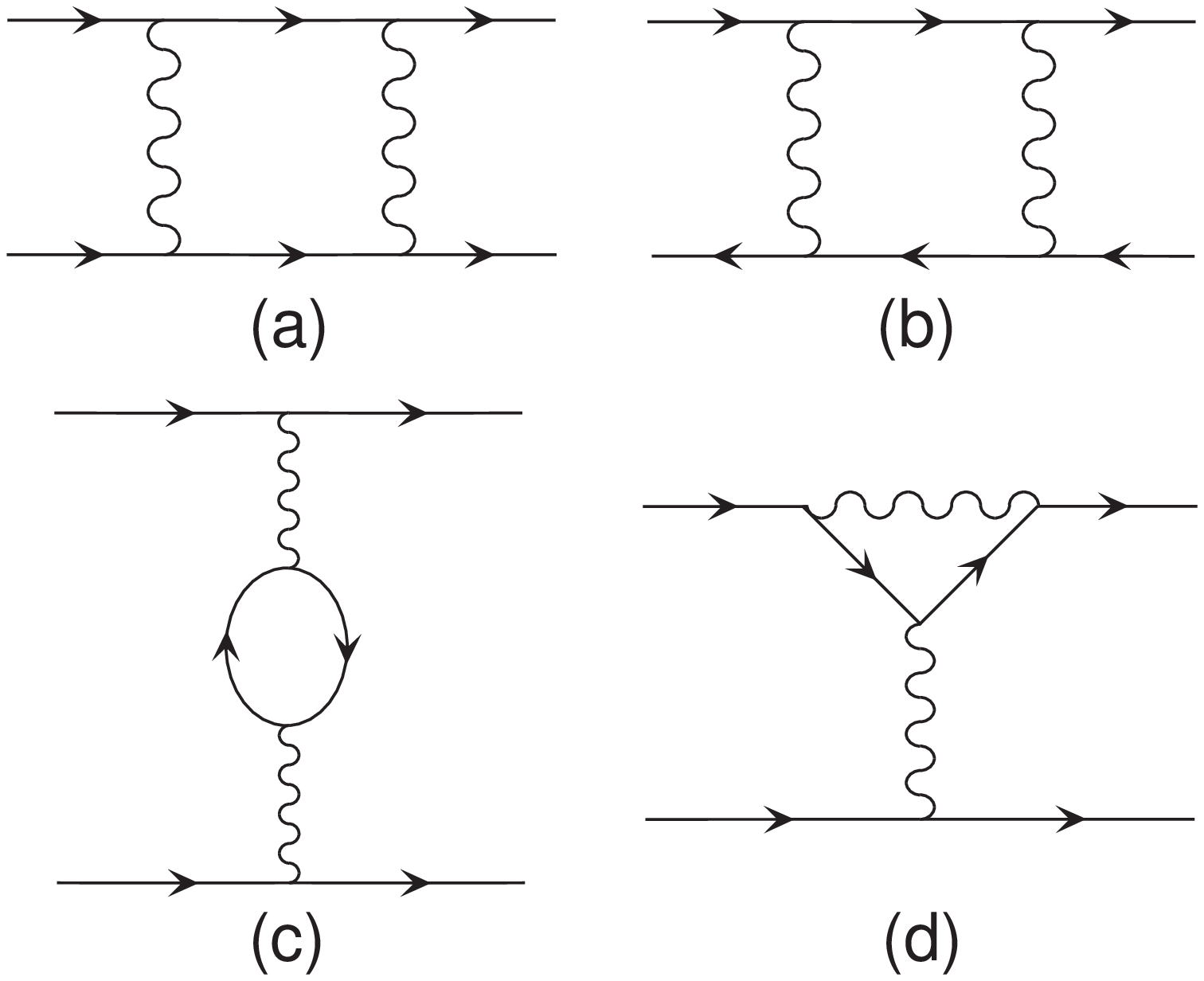}
\end{center}
\caption{Diagrams contributing to the one-loop order correction to
coupling constants.}
\label{G1loop}
\end{figure}

\begin{figure}
\begin{center}
\leavevmode\epsfysize=2cm
\epsfbox{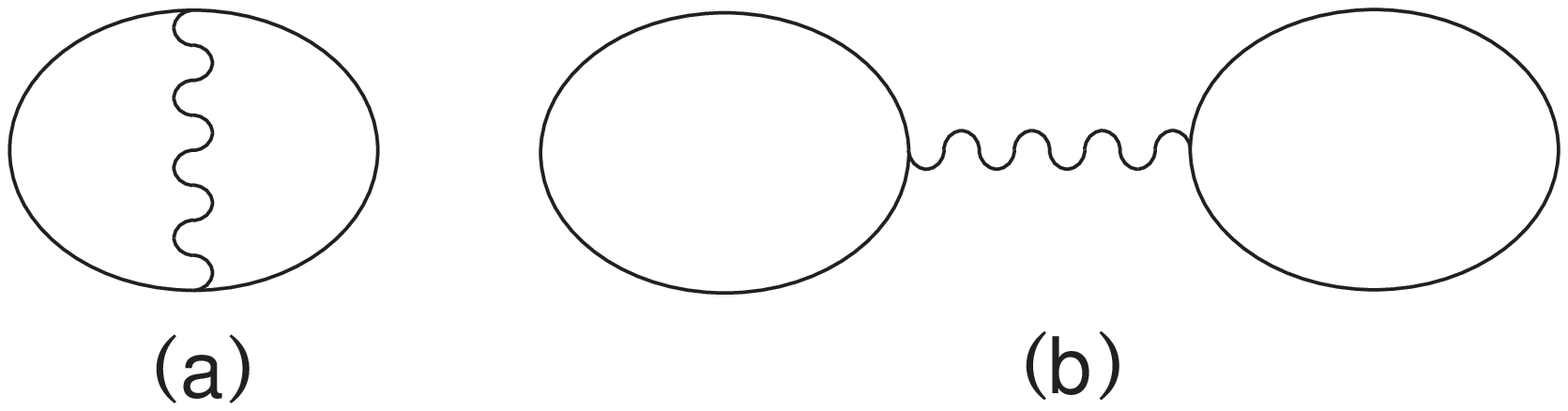}
\end{center}
\caption{Diagrams contributing to the one-loop order correction to
response functions.}
\label{R1loop}
\end{figure}

\begin{figure}
\begin{center}
\leavevmode\epsfysize=4.5cm
\epsfbox{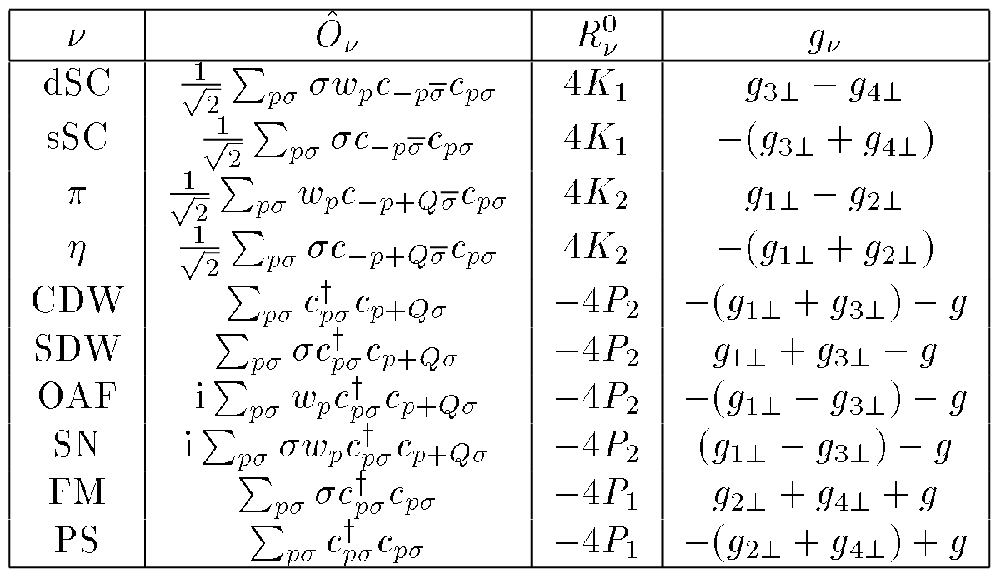}
\end{center}
\caption{Operator $\hat{O}_{\nu}$, bare response function 
$R_{\nu}^{0}(>0)$ and 
coupling constant $g_{\nu}$ for each correlation $\nu$.
Each response function can be divergent (or there exists mean-field solution
$<\hat{O}_{\nu}>\neq 0$)
for $g_{\nu}>0$. $g$ is defined as  $g\equiv g_{1\parallel}-g_{2\parallel}$.
In the RG method where only electrons near the saddle points 
are taken into account,
the sum over $p$ is restricted to $p\sim(\pi,0),\:(0,\pi)$
and the $p$-dependence of the $d_{x^{2}-y^{2}}$-wave factor 
$w_{p}\propto \cos p_{x}-\cos p_{y}$ is ignored.
In our calculation, we take $w_{p}=\sgn(\cos p_{x}-\cos p_{y})$.}
\label{gn}
\end{figure}

\subsection{Phase Diagram}

We solve the scaling equations, eq. (\ref{intflow}) and (\ref{Rnflow}),
with the initial conditions,
\bse
\bea
g^{0}_{1\perp}=g^{0}_{3\perp}&\equiv& U+2V_{Q}=U-4V,\\
g^{0}_{2\perp}=g^{0}_{4\perp}&\equiv& U+2V_{0}=U+4V,\\
g^{0}_{1\parallel}=g^{0}_{3\parallel}&\equiv& 2V_{Q}=-4V,\\
g^{0}_{2\parallel}=g^{0}_{4\parallel}&\equiv& 2V_{0}=4V,
\eea
\ese
at $x=x_{i}$. 
Here, we take $x_{i}\equiv 0$ and $R_{\nu}(x_{i})\equiv 0$ for simplicity,
although the solution of the scaling equations 
depends on the value of $x_{i}$ and $R_{\nu}(x_{i})$.

Before we show our results,
we refer to previous results for $U>0$ and $V=0$ obtained by many authors.
$H_{U}$ can be rewritten as follows,
\be
H_{U}=
\frac{U}{2}\sum_{i}n_{i}n_{i}-\frac{U}{2}\sum_{i}n_{i}.\label{nini}
\ee
If we regard the second term in the r.h.s of eq. (\ref{nini})
as the chemical potential shift,
we can take $g^{0}_{i\perp}=g^{0}_{i\parallel}=U$ as the initial conditions
and therefore use eq. (\ref{intflow2}) as the scaling equations of the
coupling constants.
For the perfect nesting case $r\equiv |t'|/t=0$,
Schulz\cite{Schulz} and { Dzyaloshinski\u\i}\cite{scvHs} 
showed that SDW occurs, 
and pointed out that small deviations from half filling lead to dSC. 
Lederer  {\em et al.}\cite{Lederer} and
Furukawa {\em et al.}\cite{FurukawaRice} solved the scaling equations 
for $r\ll 1$
and the same results.
Especially, Furukawa {\em et al.}\cite{FurukawaRice}
indicated that the correlation of $\pi$-triplet pair is suppressed to zero.
In these calculations, however,
$P_{1}$ and $K_{2}$ are neglected in eq. (\ref{intflow2})
for the reason that they are less singular than $P_{2}$, 
i.e., $c,c''\ll c'$ for $r\ll 1$, in eq. (\ref{bubbles2}) and
(\ref{coefc}).
On the other hand,
Alvarez {\em et al.}\cite{ttu} solved the flow equations 
with the initial condition $g^{0}_{i\perp}=U$ and $g^{0}_{i\parallel}=0$,
by neglecting the generation of $g_{i\parallel}$ and
omitting particle-particle diagram $K_{1}$ and $K_{2}$
in eq. (\ref{intflow}), and showed that dSC, SDW and FM can be stabilized
depending on $U/t$ and $r$.

Now, we show the phase diagram in the plane of $U>0$ and $V$
for $t'=-1/5$
in Fig.~\ref{RGpd},
in which the ordered state with highest onset temperature is shown.
First, we discuss the case that 
each ordered state is treated independently, 
because we cannot examine the coexistence of different orders 
in the RG method.
The crucial difference from our mean field result
is that superconductivity appears even for $U>0$ and $V\geq 0$, i.e.,
dSC for $U>4V$ and sSC for $U<4V$.
This result that dSC is possible for small $V>0$ as well as $V=0$ 
near half filling is consistent with a recent calculation 
based on the fluctuation-exchange (FLEX) approximation.\cite{scLRC} 
Except for superconductivity for $U>0$ and $V\geq 0$, 
the RG phase diagram is qualitatively 
same as the mean field one when we do not take the coexistence of different
orders into account, i.e.,
SDW and CDW appear for $U>4V$ and $U<4V$, respectively,
and attractive $V<0$ favors dSC for small $|V|$ and PS for large $|V|$,
respectively.
With regard to the correlation of $\pi$-triplet pair,
our RG calculation has shown that 
it can be divergent for attractive $V<0$ and large $|V|$
but is always subdominant.
Similarly, FM cannot be the most dominant solely.
These results are also consistent with our mean field ones.

Next, we discuss the possibility 
of the coexistence of different orders at low temperature,
especially at $T=0$.
It is very important that our RG calculation shows the existence of 
a region where the onset temperature of SDW or CDW
becomes highest, as shown in Fig.~\ref{RGpd}.
Since our mean field calculation in \S \ref{vnega} or \S \ref{vposi} 
shows that the Fermi surface remains in the SDW and CDW states 
near half filling,
we might expect to find a second phase transition at lower temperature
in such SDW and CDW states. 
Therefore,
at lower temperature in the SDW region in Fig.~\ref{RGpd},
(1) the coexistent state with dSC, SDW {\em and} $\pi$-triplet pair 
found for $V<0$  in the mean field approximation 
might be expected to survive 
for not only $V<0$ but also $V\geq 0$,
and (2) the ferrimagnetic coexistent state with CDW, SDW {\em and} FM
found for $U>4V>0$ in the mean field approximation 
might be expected to survive.
In fact,
as we have pointed out in \S \ref{vnega},
VMC calculations for $U>0$ and $V=0$ show the coexistence of 
dSC and SDW at low temperature,\cite{giar,yamaji}
although $\pi$-triplet pair has been neglected.
Similarly,
at lower temperature in the CDW region in Fig.~\ref{RGpd},
(3) the ferrimagnetic coexistent state with CDW, SDW {\em and} FM
found for $4V>U>0$ in the mean field approximation 
might be expected to survive.
Especially,
$\pi$-triplet pair (FM),
which cannot be stabilized solely
in the parameter region considered
in the present mean field and RG approximation, 
might be expected to arise as a result of the coexistence of 
dSC and SDW (CDW and SDW).
In order to assess the effect of 
$\pi$-triplet pair (FM) on the coexistence of dSC and SDW (CDW and SDW),
we need another theoretical treatment. 

\begin{figure}
\begin{center}
\leavevmode\epsfysize=5cm
\epsfbox{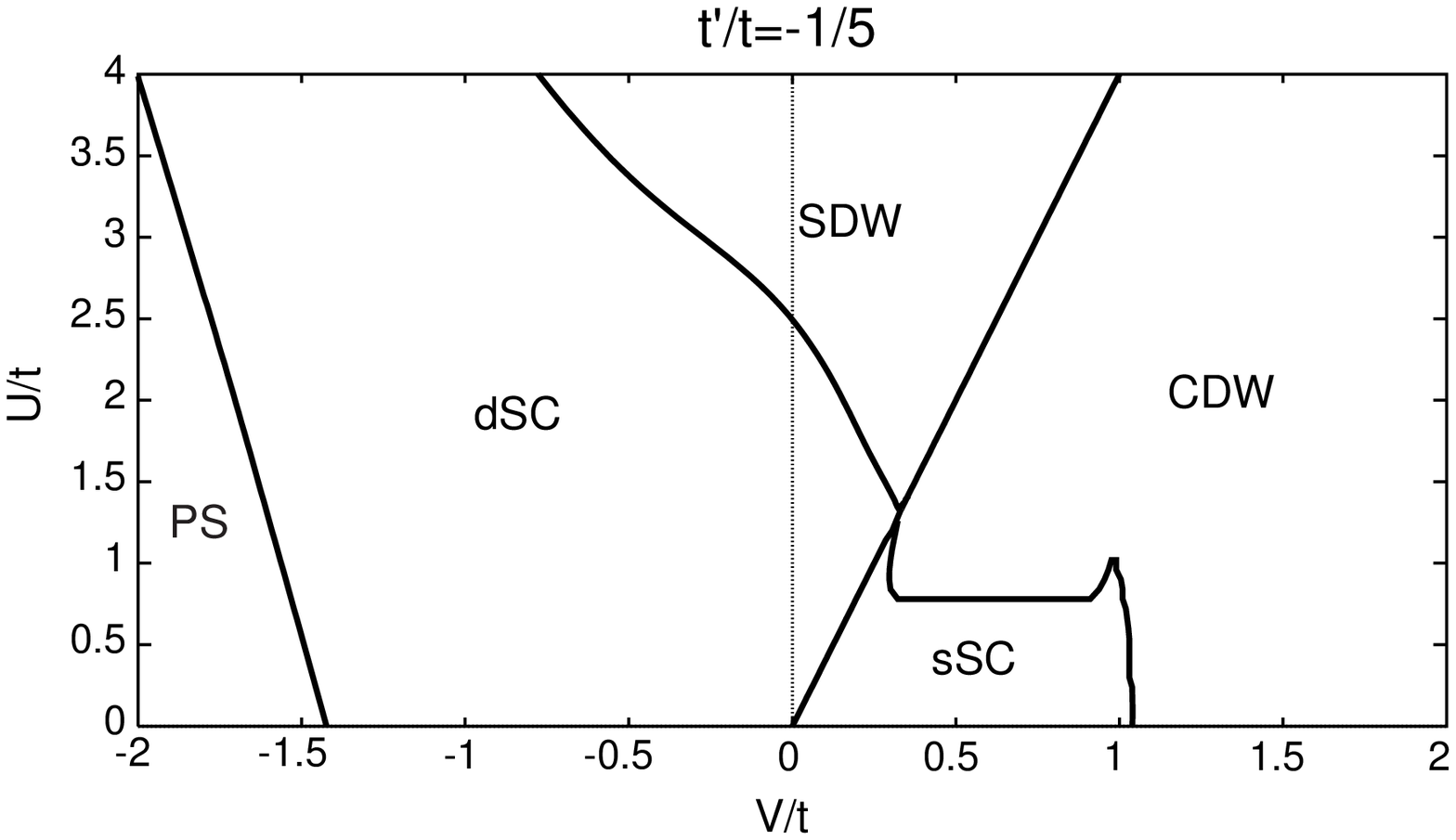}
\end{center}
\caption{The RG phase diagram for $t'/t=-1/5$ and $\mu=4t'$.}
\label{RGpd}
\end{figure}

Here, we refer to the possibility of the coexistence of sSC and CDW.
In the g-\'ology model in I,
it can be easily shown that
the coexistent state with sSC, CDW {\em and} $\eta$-singlet pair
can be stabilized near half filling at low temperature
for $g_{3\perp}<0$ and $g_{1\perp}<0$.\cite{oreD}
As seen from eq. (\ref{tolattice}),
this case corresponds to that of $U<0$ and $V>0$
in the present extended Hubbard model.
Therefore, this coexistent state might be expected to be stabilized
also in the extended Hubbard model for $U<0$ and $V>0$
in the mean field approximation.
Moreover, based on the above discussion,
it might be expected to survive in the presence of fluctuation
for not only $U<0$ but also $U\geq 0$,
at lower temperature in the CDW region in Fig.~\ref{RGpd}.
This will be reported elsewhere.

Finally, we refer to the ambiguities of the above RG method. 
Since there exist not only $\log$- but also
$\log^{2}$-divergence in the particle-particle and particle-hole 
correlation functions, eq. (\ref{bubbles2}),
we cannot safely take the limit $\omega\rightarrow 0$ 
in the scaling equations of coupling constants and
response functions, eq. (\ref{intflow}) and (\ref{Rnflow}), 
i.e., it is not clear at all whether the above RG treatment is
valid or not.
In fact, the solution of eq. (\ref{intflow}) and (\ref{Rnflow})
depends on the initial value $x_{i}$.
If we consider only the most singular $\log^{2}$ term 
in $K_{1}$ (and $P_{2}$ for $r=0$)
and take $y\equiv x^{2}=\log^{2}\frac{E_{c}}{\omega}$ 
as a new scaling variable,\cite{Schulz}
we can safely take $\omega\rightarrow 0$ limit in the scaling equations of 
coupling constants.
In this case, the above RG method
might correspond to a parquet summation of leading $\log^{2}$
divergences, rather than renormalization procedure.\cite{newVHS}

\section{Conclusion and Discussion}
We have studied in detail possible ordered states,
especially possible coexistence of different orders,
near half filling in the 2D extended Hubbard model with
on-site repulsion $U>0$ and
nearest-neighbor interaction $V$,
with emphasis on 
electrons around the saddle points 
($\pi$,0) and (0,$\pi$).

First, we have determined the phase diagram at $T=0$ 
in the mean field approximation. 
For $V<0$, we have shown 
that the coexistent state with dSC, SDW {\em and} $\pi$-triplet pair
can be stabilized near half filling.
Here, we have indicated the following important fact
which has often been neglected in previous studies:
{\em 
when we discuss the coexistence of dSC and SDW,
it is necessary to take $\pi$-triplet pair into account from the outset,
because in general the coexistence of dSC and SDW results in 
$\pi$-triplet pair
and is affected by $\pi$-triplet pair.}\cite{ore2and3}
Especially, 
$\pi$-triplet pair, which cannot condensate solely
in the present model, 
can arise through the coexistence of dSC and SDW.
Since the phase separation (PS) is generally expected to occur
in the presence of finite-range attractive interaction such as $V<0$, 
we have examined the effect of PS on the mean field ground state 
in the random phase approximation (RPA).
The coexistent state with dSC, SDW and $\pi$-triplet pair
is severely hampered by PS but survives,
and that the long-range Coulomb repulsion such as
next-nearest-neighbor density-density repulsion
leads to the suppression of PS. 
On the other hand, for $V>0$, we showed that
a {\em ferrimagnetic} coexistent state with
commensurate charge-density-wave (CDW), SDW {\em and} ferromagnetism (FM)
can be stabilized near half filling.
Here, we have indicated the following important fact:
{\em 
when we discuss the coexistence of CDW and SDW,
it is necessary to take FM into account from the outset,
because in general the coexistence of CDW and SDW results in FM
and is affected by FM.}
Especially, FM, which cannot be stabilized solely
with the present choice of parameters,
can arise through the coexistence of CDW and SDW.
It is to be noted that 
the above mean field coexistent states near half filling
can be stabilized at low temperature, especially at $T=0$, 
when CDW or SDW, in which the Fermi surface remains near half filling, 
arises first at high temperature.


In order to examine the effects of fluctuation on
the mean field ordered states, 
we have adopted the RG method
for the {\em special} case that 
the Fermi level lies just on the saddle points.
We have shown that 
the crucial difference from our mean field result is
that superconductivity can arise even for $U>0$ and $V\geq0$;
dSC and sSC for $U>4V$ and $U<4V$, respectively.
Except for this difference,
the RG phase diagram is qualitatively 
same as the mean field one when we do not take the coexistence of different
orders into account, {\em e.g.}, 
SDW and CDW can arise for $U>4V$ and $U<4V$, respectively.
Especially, the correlation of $\pi$-triplet pair or FM
cannot be the most dominant solely.
Here, it is very important that 
a region where the onset temperature of SDW or CDW
becomes highest is found in the RG phase diagram.
Since the Fermi surface remains near half filling 
in these SDW and CDW states,
we might expect to find a second phase transition at lower temperature.
In the RG method, however,
we cannot assess such possibilities.
On the other hand,
the mean field approximation,
which is often questionable for the 2D case,
is of great advantage in that 
we can study the stability of coexistent states with
different order parameters quantitatively. 
Therefore,
we can conclude that our mean field calculations 
indicate the possibilities that
(1) SDW, which has been shown in our RG calculation 
to arise first at high temperature for $U>4V$,  
coexists with dSC {\em and} $\pi$-triplet pair, or with
CDW {\em and} FM, at lower temperature, and that
(2) CDW, which has been shown in our RG calculation to arise
first at high temperature for $U<4V$,  
coexists with SDW {\em and} FM, at lower temperature.


Throughout this letter, 
we have assumed YBCO-type Fermi surface by introducing $t'$ and
consider only {\em commensurate} (C) SDW or CDW.
Near half filling, however, {\em incommensurate} (IC) ordering 
or stripe formation can be expected, especially for $t'=0$ 
in the repulsive Hubbard model.\cite{schulzIC,riceIC,machida} 
Recently, the effect of $V$ on such stripe states 
has been examined.\cite{UVIC}
The effect of IC ordering on the stability of
the coexistent states
(with dSC, C-SDW and $\pi$-triplet pair, 
and with C-CDW, C-SDW and FM) is beyond the scope of the present study.
In the repulsive Hubbard model with $t'=0$, 
Giamarchi {\em et al.}\cite{giarICHU} have shown that
the coexistent state with dSC and C-SDW state
have higher energy than that with dSC and IC-SDW.

With regard to dSC,
we have assumed that the superconducitng gap symmetry 
is purely of $d_{x^{2}-y^{2}}$-wave.
However, there are a few indications that 
dSC mixed with components of other symmetry
can be stabilized,\cite{onlyV,laughlin,ogata,kuroki}
dependent on interaction, electron density, etc.
Such mixed pairing states leave much room for future studies.


The author thanks to H. Fukuyama, H. Kohno and M. Ogata 
for valuable discussions.

\appendix
\section{Symmetry among Various Order Parameters}\label{sym}
We examine the relationship among the order parameters 
as shown in Fig.~\ref{gn}.
First,
the operators of dSC, SDW and $\pi$-triplet pair are 
equivalent in that they are 'rotated' into each other,
\bse
\be
[\hat{O}_{SDW},\hat{O}_{dSC}]= 2\hat{O}_{\pi},\:\:
[\hat{O}^{\dagger}_{\pi},\hat{O}_{SDW}]=2\hat{O}^{\dagger}_{dSC},
\ee
\be
[\hat{O}_{dSC},\hat{O}^{\dagger}_{\pi}]=2\hat{O}_{SDW}.\label{so5app1}
\ee
\label{so5rot1}
\ese
This relationship underlies the $SO(5)$ theory,
in which 
dSC and SDW are unified into a {\em five}-dimensional vector
{\em superspin} and
$\hat{O}_{\pi}$ describes excitation towards the SDW (dSC) direction
in the dSC (SDW) ground state.\cite{SCZhang}
We note that
eq. (\ref{so5app1}) holds due to $w_{p}^{2}=1$,
i.e., $w_{p}\equiv\sgn (\cos p_{x}-\cos p_{y})$.
If $w_{p}\propto \cos p_{x}-\cos p_{y}$, 
it is satisfied only in the long wavelength limit where
only electrons near the saddle points are important.
It is very important to note that 
if two of the above three order parameters coexist,
another one also results generally.
As we have already shown in the mean field approximation,
the coexistent state with dSC, SDW {\em and} $\pi$-triplet pair
can be stabilized near half filling in the 2D extended Hubbard model
for $U>0$ and $V<0$.

Next, the operators of sSC, CDW and $\eta$-singlet pair 
are equivalent in that they are 'rotated' into each other,
\bse
\be
[\hat{O}_{CDW},\hat{O}^{\dagger}_{sSC}]= 2\hat{O}^{\dagger}_{\eta},\:\:
[\hat{O}_{\eta},\hat{O}_{CDW}]=2\hat{O}_{sSC},\\
\ee
\be
[\hat{O}^{\dagger}_{sSC},\hat{O}_{\eta}]=2\hat{O}_{CDW},
\ee
\label{so3rot1}
\ese
This relationship underlies the $SO(3)$ theory,
in which sSC and CDW are unified into a {\em three}-dimensional
vector {\em pseudospin}
and $\hat{O}_{\eta}$ describes excitation towards the CDW (sSC) direction
in the sSC (CDW) ground state.\cite{Eta1,Eta2}
In fact, such properties are useful in the {\em attractive} Hubbard model.
It is very important to note that 
if two of the above three order parameters coexist,
another one also results generally.
It can be easily shown in the mean field approximation that 
a state where sSC, CDW {\em and} $\eta$-singlet pair coexist
can be stabilized near half filling in the g-\'ology model used in I
for $g_{3\perp}<0$ and $g_{1\perp}<0$.\cite{oreD}

We note that similar relationships hold
among dSC, OAF and $\eta$-singlet pair,
\bse
\be
[\hat{O}^{\dagger}_{\eta},\hat{O}_{dSC}]=2{\rm i}\hat{O}_{OAF},\:\:
[\hat{O}_{OAF},\hat{O}_{\eta}]=2{\rm i}\hat{O}_{dSC},
\ee
\be
[\hat{O}^{\dagger}_{dSC},\hat{O}_{OAF}]= 2{\rm i}\hat{O}^{\dagger}_{\eta},
\label{so3app}
\ee
\label{so3rot2}
\ese
and among sSC, SN and $\pi$-triplet pair,
\bse
\be
[\hat{O}^{\dagger}_{sSC},\hat{O}_{\pi}]=2{\rm i}\hat{O}_{SN},\:\:
[\hat{O}_{sN},\hat{O}_{sSC}]= 2{\rm i}\hat{O}_{\pi},
\ee
\be
[\hat{O}^{\dagger}_{\pi},\hat{O}_{SN}]=2{\rm i}\hat{O}^{\dagger}_{sSC}.
\label{so5app2}
\ee
\label{so5rot2}
\ese
If $w_{p}\propto \cos p_{x}-\cos p_{y}$, 
eq. (\ref{so3app}) and (\ref{so5app2}) hold only approximately.
In each case,
if two of the above three order parameters coexist,
another one also results generally.
It can be easily shown in the mean field approximation that 
a state where
dSC, OAF {\em and} $\eta$-singlet pair 
(sSC, SN {\em and} $\pi$-triplet pair) coexist 
can be stabilized in the g-\'ology model used in I for 
$g_{3\perp}>0$ and $g_{1\perp}<0$  
($g_{3\perp}<0$ and $g_{1\perp}>0$).\cite{oreD}

Moreover, eq. (\ref{so5rot1})-(\ref{so5rot2}) imply close relationships 
(1) between sSC and dSC, (2) between $\eta$-singlet pair
and $\pi$-triplet pair, and
(3) between density-wave (CDW, SDW) and local-current (OAF, SN).
In fact,
the {\em hermite} operators of uniform bond-charge and bond-spin 
(BC and BS) density 
with $d_{x^{2}-y^{2}}$-symmetry defined as 
\be
\hat{O}_{BC}=\sum_{p\sigma}w_{p}c^{\dagger}_{p\sigma}c_{p\sigma},\:\:
\hat{O}_{BS}=\sum_{p\sigma}\sigma w_{p}c^{\dagger}_{p\sigma}c_{p\sigma},
\ee
satisfy similar commutation relations,
\bse
\be
[\hat{O}^{\dagger}_{\alpha},\hat{O}_{\beta}]=2\hat{O}_{\gamma},\:\:
[\hat{O}_{\gamma},\hat{O}^{\dagger}_{\alpha}]=2\hat{O}^{\dagger}_{\beta},
\ee
\be
[\hat{O}_{\beta},\hat{O}_{\gamma}]=2\hat{O}_{\alpha},\label{appCOPI}
\ee
\ese
for $(\alpha,\beta,\gamma)=$(sSC, dSC, BC) and ($\eta$, $\pi$, BS),
and
\bse
\be
[\hat{O}_{\alpha},\hat{O}_{\beta}]=2{\rm i}\hat{O}_{\gamma},\:\:
[\hat{O}_{\gamma},\hat{O}_{\alpha}]=2{\rm i}\hat{O}_{\beta},
\ee
\be
[\hat{O}_{\beta},\hat{O}_{\gamma}]=2{\rm i}\hat{O}_{\alpha},\label{appDWLC}
\ee
\ese
for 
$(\alpha,\beta,\gamma)=$(SDW, BC, SN), (CDW, BC, OAF), 
(SDW, BS, OAF) and (CDW, BS, SN).
If  $w_{p}\propto \cos p_{x}-\cos p_{y}$, 
eq. (\ref{appCOPI}) and (\ref{appDWLC}) hold only approximately in each case.

The relationship among order parameters 
which are rotated into each other by
$\hat{O}_{\eta}$, $\hat{O}_{\pi}$, $\hat{O}_{BC}$ and $\hat{O}_{BS}$
is summarized as shown in Fig.~\ref{symmetry},
where the connection with $\hat{O}_{dSC}$ and $\hat{O}_{sSC}$ by 
$\hat{O}_{BC}$ is not explicitly shown.

\begin{figure}
\begin{center}
\leavevmode\epsfysize=5cm
\epsfbox{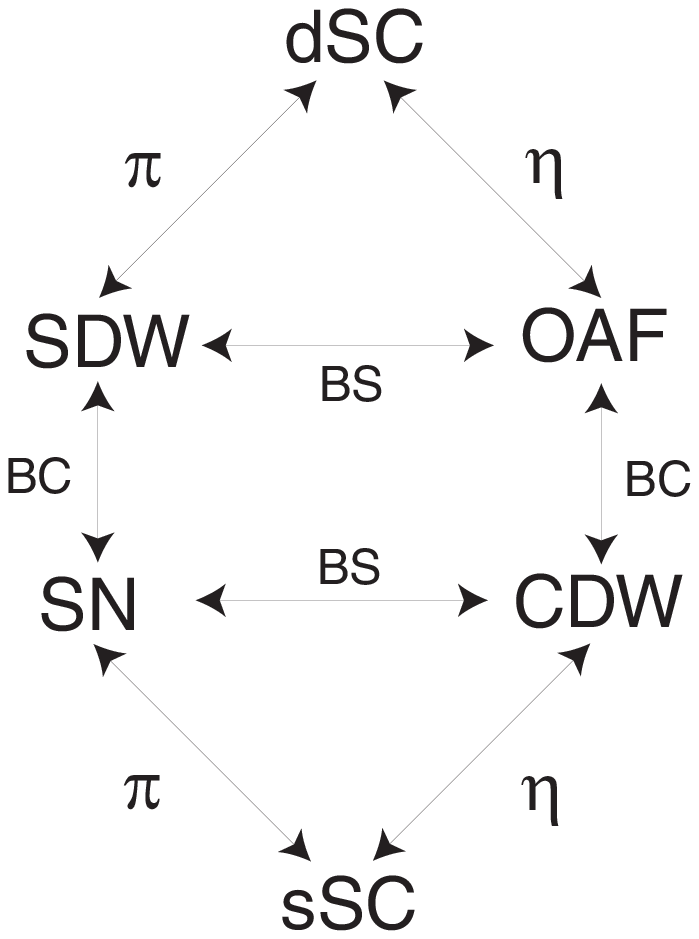}
\end{center}
\caption{The relationship among various order parameters.}
\label{symmetry}
\end{figure}

\section{Charge Susceptibility in the Random Phase Approximation}\label{RPA}

The charge compressibility $\kappa$
is equal to static and uniform 
charge susceptibility $\chi$,
\begin{equation}
\kappa=\lim_{q\rightarrow 0}\chi^{R}(q,q,\omega=0),\label{kappadef}
\end{equation}
where $\chi^{R}(q,q',\omega)$ is the retarded density-density correlation 
function, which is obtained from the analytic continuation of 
$\chi (q,q',{\rm i}\omega_{l})$ through 
${\rm i}\omega_{l}\rightarrow\omega+{\rm i}\delta$,
\begin{subequations}
\begin{eqnarray}
\chi (q,q',{\rm i}\omega_{l})&=&
\int_{0}^{\beta}{\rm d}\tau{\rm e}^{{\rm i}\omega_{l}\tau}
\chi (q,q',\tau).
\end{eqnarray}
\end{subequations}	
$\chi (q,q',\tau)$ is the two-particle thermal Green function 
in imaginary time given by
\begin{subequations}
\begin{eqnarray}
\chi(q,q',\tau)&=&\sum_{\sigma\sigma '}
N_{\sigma\sigma '}(q,q',\tau),\\
N_{\sigma\sigma '}(q,q',\tau)
&=&\frac{1}{N}<T_{\tau}n_{q\sigma}(\tau)n_{-q'\sigma '}>,
\label{eqnnab}
\end{eqnarray}
\end{subequations}	
where
$n_{q\sigma}=\sum_{k}c^{\dagger}_{k\sigma}c_{k+q\sigma}$,
$n_{-q\sigma}=(n_{q\sigma})^{\dagger}$.
We write $N_{\sigma\sigma '}$ in the matrix form as follows,
\be
\hat{N}=
\left (
\begin{array}{cc}
N_{\uparrow\uparrow}&N_{\uparrow\downarrow}\\
N_{\downarrow\uparrow}&N_{\downarrow\downarrow}
\end{array}
\right ).
\ee

For the mean field Hamiltonian 
with $\Delta_{dSC}$, $\Delta_{SDW}$ and $\Delta_{\pi}$,
the one-particle thermal Green functions, defined by
\begin{subequations}
\begin{eqnarray}
G_{\sigma\sigma '}(p,p',\tau)&\equiv&
-<T_{\tau}c_{p\sigma}(\tau)c^{\dagger}_{p'\sigma '}>,\\
F^{\dagger}_{\sigma\sigma '}(p,p',\tau)&\equiv&
-<T_{\tau}c^{\dagger}_{p\sigma}(\tau)c^{\dagger}_{p'\sigma '}>,\\
F_{\sigma\sigma '}(p,p',\tau)&\equiv&
-<T_{\tau}c_{p\sigma}(\tau)c_{p'\sigma '}>,
\end{eqnarray}
\end{subequations}
have the following form,
\begin{subequations}
\begin{eqnarray}
G_{\sigma\sigma}(p,p',\tau)&\equiv&
\delta_{p',p}G_{\sigma\sigma 1}(p,\tau)
+\delta_{p',p+Q}G_{\sigma\sigma 2}(p,\tau),\makebox[3em]{}\\
F^{(\dagger)}_{\sigma\overline{\sigma}}(p,p',\tau)&\equiv&
\delta_{p',-p}F^{(\dagger)}_{\sigma\overline{\sigma}1}
(p,\tau)+\delta_{p',-p+Q}F^{(\dagger)}_{\sigma\overline{\sigma}2}(p,\tau).
\makebox[3em]{}
\end{eqnarray}
\end{subequations}
Therefore, it is seen that $\hat{\chi}$ has the following form,
\bse
\be
\hat{N}(q,q',\tau)=
\delta_{q',q}\hat{N}_{1}(q,\tau)+
\delta_{q',q+Q}\hat{N}_{2}(q,\tau),
\ee
\be 
\hat{N}_{1}=\left (
\begin{array}{cc}
N_{1\parallel}& N_{1\perp}\\
N_{1\perp}& N_{1\parallel}\\
\end{array}
\right),
\ee
\be
\hat{N}_{2}=\left (
\begin{array}{cc}
N_{2\parallel}& N_{2\perp}\\
-N_{2\perp}& -N_{2\parallel}\\
\end{array}
\right),
\ee
\label{Nform}
\ese
The subscript 1 and 2 represent normal and Umklapp part,
respectively. It is to be noted that there exists Umklapp part $\hat{N}_{2}$
for $\Delta_{SDW}\neq 0$ or $\Delta_{\pi}\neq 0$.
With regard to $\chi(q,q',\tau)$, it has only normal part,
\bse
\bea
\chi(q,q',\tau)&=&\delta_{q',q}\cdot \chi_{0}(q,\tau),\\
\chi_{0}(q,\tau)&\equiv&2\left\{ 
N_{1\parallel}(q,\tau)+N_{1\perp}(q,\tau)\right\}.\label{chi0}
\eea
\ese

Next, we treat the effect of interaction in the 
random phase approximation (RPA).
The RPA equation is diagrammatically shown in Fig.~\ref{RPAeqn}.
Here, we consider only (a) for interaction vertex 
in Fig.~\ref{RPAeqn}, which leads to the diagram
shown in Fig.~\ref{RPAdiagram1}.
In the frequency space, the RPA equation is written as follows 
($z\equiv {\rm i}\omega_{l}$),
\bse
\bea
\hat{N}_{RPA}(q,q',z)=\hat{N}(q,q',z)\makebox[8em]{}&&\nonumber\\
+\sum_{q_{1}}
\hat{N}(q,q_{1},z)\hat{g}(q_{1})\hat{N}_{RPA}(q_{1},q',z),
\makebox[1em]{}&&
\eea
where 
\be
\hat{g}
\equiv\left (
\begin{array}{cc}
g_{\parallel}& g_{\perp}\\
g_{\perp}& g_{\parallel}
\end{array}
\right ),
\end{equation}
\be
g_{\parallel}(q)\equiv -2V_{q},\:\:
g_{\perp}(q)\equiv -(U+2V_{q}),
\ee
\ese
and $V_{q}=V(\cos q_{x}+\cos q_{y})$.
Since it can be seen that 
$\hat{N}_{RPA}$ has the same form as eq. (\ref{Nform}),
$\chi_{RPA}$ is obtained as follows,
\bse
\bea
\chi_{RPA}(q,q)&=&
2\frac{n^{+}_{q}X^{-}_{q+Q}+Y_{q}}
{X^{+}_{q}X^{-}_{q+Q}-g^{+}_{q}Y_{q}},\makebox[3em]{}\label{kappaRPA}\\
\chi_{RPA}(q,q+Q)&\equiv& 0,
\eea
where 
\be
n^{\pm}_{q}\equiv N_{1\parallel}(q)\pm N_{1\perp}(q),\:\:
u^{\pm}_{q}\equiv N_{2\parallel}(q)\pm N_{2\perp}(q),
\ee
\be
g^{\pm}_{q}\equiv g_{\parallel}(q)\pm g_{\perp}(q),
\ee
\be
X^{\pm}_{q}=1-g^{\pm}_{q}n^{\pm}_{q},\:\:
Y_{q}=g^{-}_{q+Q}u^{-}_{q}u^{+}_{q+Q},
\ee
\ese
and the variable $z$ is not explicitly written.
Therefore, 
we can obtain the charge compressibility $\kappa$ in the RPA, 
from eq. (\ref{kappadef}) and (\ref{kappaRPA}).
In the normal or dSC state,
$\chi_{RPA}$ has a familiar form due to $u_{q}^{\pm}\equiv 0$,
\be
\chi_{RPA}(q,q)=\frac{2 n_{q}^{+}}{X^{+}_{q}}
=\frac{\chi_{0}(q)}{1+\frac{1}{2}(U+4V_{q})\chi_{0}(q)},
\ee
where $\chi_{0}$ is given by eq. (\ref{chi0}).
However, since $u_{q}^{\pm}\neq 0$ 
in the SDW or $\pi$-triplet pairing state,
especially in the coexistent state with dSC, SDW and $\pi$-triplet pair,
$\chi_{RPA}$ has a nontrivial form.

\begin{figure}
\begin{center}
\leavevmode\epsfysize=5cm
\epsfbox{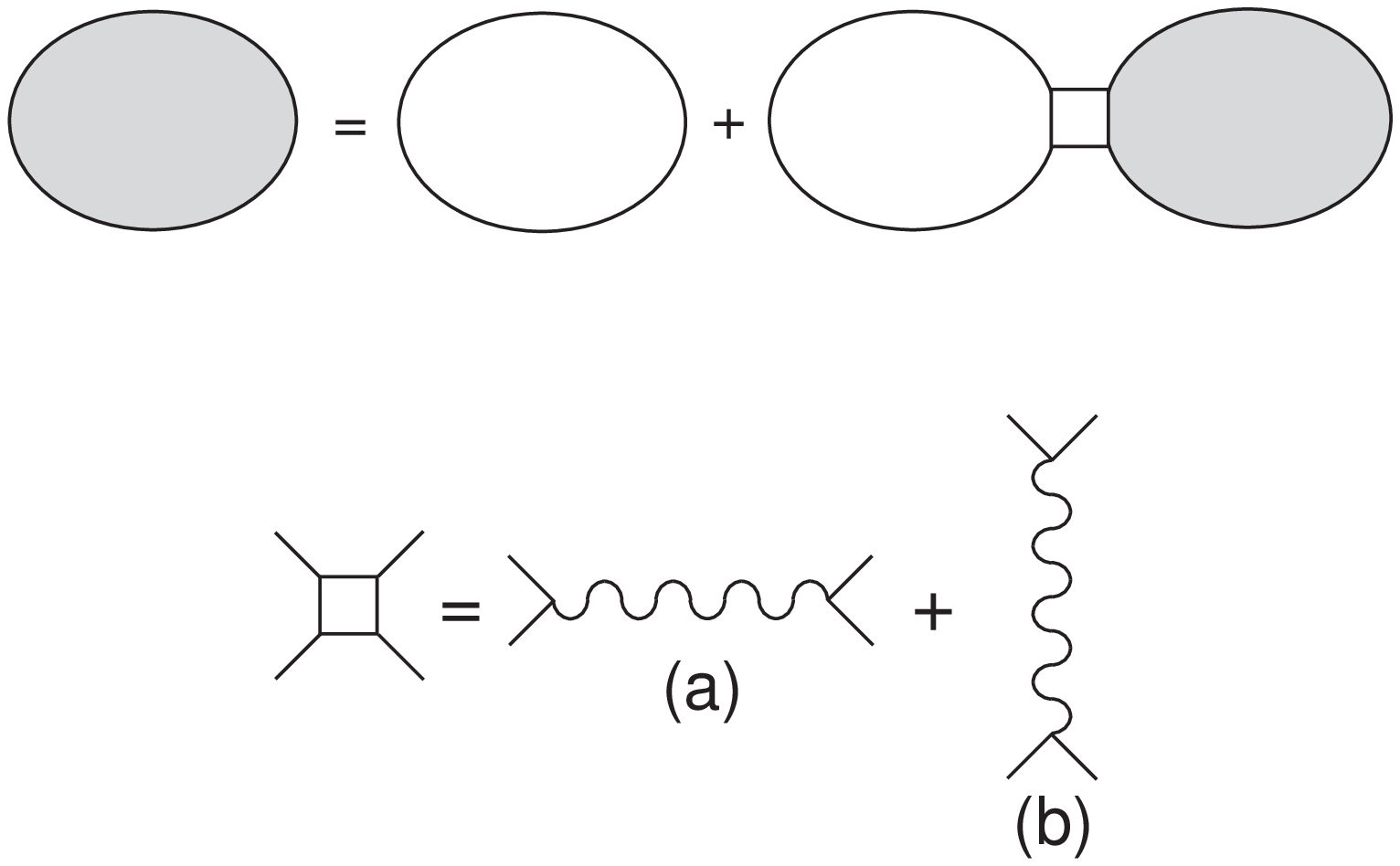}
\end{center}
\caption{The RPA equation for charge susceptibility.
The square stands for interaction vertex.
In our calculation, the diagram of (b) 
for interaction vertex is neglected.}
\label{RPAeqn}
\end{figure}


\begin{thebibliography}{10}

\bibitem{ore2and3}
M. Murakami and H. Fukuyama: J. Phys. Soc. Jpn. {\bf 67} (1998) 41; 2784.

\bibitem{EHM2D}
R. Micnas, J. Ranninger, S. Robaszkiewicz and S. Tabor: Phys. Rev. B {\bf 37}
  (1988) 9410; R. Micnas, J. Ranninger and S. Robaszkiewicz: Phys. Rev. B {\bf
  39} (1989) 11653; R. Micnas, J. Ranninger and S. Robaszkiewicz: Rev. Mod.
  Phys. {\bf 62} (1990) 114.

\bibitem{dag}
E. Dagotto, J. Riera, Y. C. Chen, A. Moreo, A. Nazarenko, F. Alcaraz and F.
  Ortolani: Phys. Rev. B {\bf 49} (1994) 3548; A. Nazarenko, A. Moreo, E.
  Dagotto and J. Riera: Phys. Rev. B {\bf 54} (1996) R768.

\bibitem{sdd}
J. Ferrer, M. A. Gonz\'alez-Alvarez and J. S\'anchez-Ca\~nizares: Phys. Rev. B
  {\bf 57} (1998) 7470.

\bibitem{onlyV}
J. F. Annett and J. P. Wallington: condmat/9807220.

\bibitem{oreD}
M. Murakami: Ph.D. thesis (1998).

\bibitem{DZ}
E. Demler and S. C. Zhang: Phys. Rev. Lett. {\bf 75} (1995) 4126.

\bibitem{SCZhang}
S. C. Zhang: Science {\bf 275} (1997) 1089.

\bibitem{chmAAcoe}
L. Arrachea and A. A. Aligia: Physica C {\bf 303} (1998) 141.

\bibitem{crete}
G. C. Psaltakis and E. W. Fenton: J. Phys. C {\bf 16} (1983) 3913.

\bibitem{inaba}
M. Inaba, H. Matsukawa, M. Saitoh and H. Fukuyama: Physica C {\bf 257} (1996)
  299.

\bibitem{Chen}
G. J. Chen, R. Joynt, F. C. Zhang and C. Gros: Phys. Rev. B {\bf 42} (1990)
  2662.

\bibitem{giar}
T. Giamarchi and C. Lhuillier: Phys. Rev. B {\bf 43} (1991) 12943.

\bibitem{Himeda}
A. Himeda and M. Ogata: Phys. Rev. B {\bf 60} (1999) R9935.

\bibitem{yamaji}
K. Yamaji, T. Yanagisawa, T. Nakanishi and S. Koike: Physica C {\bf 304} (1998)
  225.

\bibitem{BCR1D}
S. Kivelson, W. -P. Su, J. R. Schrieffer and A. J. Heeger: Phys. Rev. Lett.
  {\bf 58} (1987) 1899.

\bibitem{BCRHS}
J. E. Hirsch: Physica C {\bf 158} (1989) 326.

\bibitem{PH1D}
D. K. Campbell, J. Tinka Gammel and E. Y. Loh, Jr.: Phys. Rev. B {\bf 42}
  (1990) 475.

\bibitem{BCAAL}
F. Marsiglio and J. E. Hirsch: Phys. Rev. B {\bf 49} (1994) 1366.

\bibitem{fluxphase}
I. Affleck and J. B. Marston: Phys. Rev. B {\bf 37} (1988) 3774.

\bibitem{fsinstability}
H. J. Schulz: Phys. Rev. B {\bf 39} (1988) 2940.

\bibitem{OAF}
A. A. Nersesyan and G. E. Vachnadze: J. Low Temp. Phys. {\bf 77} (1989) 293.

\bibitem{SN}
A. A. Nersesyan, G. I. Japaridze and I. G. Kimeridze: J. Phys. Condens. Matter
  {\bf 3} (1991) 3353.

\bibitem{CG}
B. Chattopadhyay and D. M. Gaitonde: Phys. Rev. B {\bf 55} (1997) 15364.

\bibitem{scvHs}
I. E. Dzyaloshinski\u\i: Zh. Eksp. Teor. Fiz. {\bf 93} (1987) 1487 [Sov. Phys.
  JETP {\bf 66} (1987) 848].

\bibitem{1D2kf}
N. Kobayashi, M. Ogata and K. Yonemitsu: J. Phys. Soc. Jpn. {\bf 67} (1998)
  1098.

\bibitem{penn}
D. R. Penn: Phys. Rev. {\bf 142} (1966) 350.

\bibitem{Schulz}
H. J. Schulz: Europhys. Lett. {\bf 4} (1987) 609.

\bibitem{Lederer}
P. Lederer, G. Montambaux and D. Poilblanc: J. Physique {\bf 48} (1987) 1613.

\bibitem{vHsRG}
J. Gonz\'alez, F. Guinea and M. A. H. Vozmediano: Europhys. Lett. {\bf 34}
  (1996) 711; Nucl. Phys. B {\bf 485} (1997) 694.

\bibitem{ttu}
J. V. Alvarez, J. Gonz\'alez, F. Guinea and M. A. H. Vozmediano: J. Phys. Soc.
  Jpn. {\bf 67} (1998) 1868; condmat/9804153.

\bibitem{ttu2}
J. V. Alvarez and J. Gonz\'alez: condmat/9803131.

\bibitem{FurukawaRice}
N. Furukawa, T. M. Rice and M. Salmhofer: Phys. Rev. Lett. {\bf 81} (1998)
  3195.

\bibitem{Eta1}
C. N. Yang: Phys. Rev. Lett. {\bf 63} (1989) 2144.

\bibitem{Eta2}
S. C. Zhang: Phys. Rev. Lett. {\bf 65} (1990) 120; Int. J. Mod. Phys. B {\bf 5}
  (1991) 153.

\bibitem{scLRC}
G. Esirgen, H. -B. {Sch\"uttler} and N. E. Bickers: Phys. Rev. Lett. {\bf 82}
  (1999) 1217.

\bibitem{newVHS}
J. Gonz\'alez, F. Guinea and M. A. H. Vozmediano: condmat/9905166.

\bibitem{schulzIC}
H. J. Schulz: Phys. Rev. Lett. {\bf 64} (1990) 1445.

\bibitem{riceIC}
D. Poilblanc and T. M. Rice: Phys. Rev. B {\bf 39} (1989) 9749.

\bibitem{machida}
M. Kato, K. Machida, H. Nakanishi and M. Fujita: J. Phys. Soc. Jpn. {\bf 59}
  (1990) 1047.

\bibitem{UVIC}
G. Seibold, C. Castellani, C. Di. Castro and M. Grilli: Phys. Rev. B {\bf 58}
  (1998) 13506.

\bibitem{giarICHU}
T. Giamarchi and C. Lhuillier: Phys. Rev. B {\bf 42} (1990) 10641.

\bibitem{laughlin}
R. B. Laughlin: Phys. Rev. Lett. {\bf 80} (1998) 5188.

\bibitem{ogata}
M. Ogata: J. Phys. Soc. Jpn. {\bf 66} (1997) 3375.

\bibitem{kuroki}
K. Kuroki and H. Aoki: J. Phys. Soc. Jpn. {\bf 67} (1998) 1533.

\end{thebibliography}
\end{document}